\documentclass[a4paper,12pt]{article}

\usepackage{geometry}
\usepackage[pdftex]{graphicx}
\usepackage{setspace} 

\usepackage{empheq} 
\usepackage{comment} 
\usepackage{multicol} 
\usepackage{enumitem} 
\usepackage[justification=centering]{caption}

\usepackage[table]{xcolor} 
\usepackage{multirow} 
\usepackage{float}
\usepackage{verbatim}
\usepackage{url}
\usepackage{appendix}
\usepackage{ulem}
\usepackage[hang]{footmisc}

\usepackage[english]{babel}
\usepackage{textcomp}
\usepackage{amssymb}

\usepackage{booktabs}
\usepackage{adjustbox}

\usepackage{rotating}
\usepackage{tablefootnote}
\usepackage[flushleft]{threeparttable}
\usepackage{tabularx}

\usepackage{titlesec}
\usepackage{xr-hyper}

\usepackage{pdflscape}
\usepackage{rotating}
\usepackage{anysize}
\usepackage{titlesec}

\marginsize{2.0 cm}{2.0 cm}{2.0 cm}{2.0 cm}

\usepackage[breaklinks=true,colorlinks=true,citecolor=blue]{hyperref}
\usepackage{multirow}

\usepackage{har2nat}
\bibpunct{(}{)}{;}{a}{,}{,}

\begin{document}
	
	\pagestyle{plain}

\title{\textsc{Will the last be the first? School closures and educational outcomes}}

\author{\textbf{Michele Battisti}\thanks{Department of Law, University of Palermo, Piazza Bologni 8, 90134 Palermo, Italy, e-mail: michele.battisti@unipa.it.}\\  University of Palermo, ICEA 
        \and 
        \textbf{Giuseppe Maggio}\thanks{Department of Law, University of Palermo, Piazza Bologni 8, 90134 Palermo, Italy, e-mail: giuseppe.maggio@unipa.it.}\\  University of Palermo, ICEA
        }
 \maketitle

\begin{abstract}

Governments have implemented school closures and online learning as one of the main tools to reduce the spread of Covid-19. Despite the potential benefits in terms of reduction of cases, the educational costs of these policies may be dramatic. This work identifies the educational costs, expressed as decrease in test scores, for the whole universe of Italian students attending the 5th, 8th and 13th grade of the school cycle during the 2021/22 school year. The analysis relies on a difference-in-difference model in relative time, where the control group is the closest generation before the Covid-19 pandemic. The results suggest a national average loss between 1.6-4.1\% and 0.5-2.4\% in Mathematics and Italian test scores, respectively. After collecting the precise number of days of school closures for the universe of students in Sicily, we estimate that 30 additional days of closure decrease the test score by 1\%. However, the impact is much larger for students from high schools (1.8\%) compared to students from low and middle schools (0.5\%). This is likely explained by the lower relevance of parental inputs and higher reliance on peers inputs, within the educational production function, for higher grades. Findings are also heterogeneous across class size and parental job conditions, pointing towards potential growing inequalities driven by the lack of in front teaching.

\end{abstract}

\vspace{3 cm}


\pagenumbering{arabic}
\newpage \doublespacing \setcounter{page}{2}

\section{Introduction\label{se:intro}}

After public schooling became the norm for the western societies, scholars have often referred to the schools as great equalizers, mainly because of their potential role in reducing disparities and providing similar learning to children from different socio-economic environment \citep{cremin1951american,agostinelli2022great}. A wide strand of literature has criticized this concept and suggested how students with initial advantages often attend schools with higher resources, more compelling programs and highly interactive teachers \citep{condron2003disparities,downey2004schools,roscigno2006education}. In contrast, disadvantaged students, such as the ones coming from inner cities or rural areas of the United States, are often associated to lower educational achievements and higher likelihood of dropping out during high-schools \citep{roscigno2006education}. Whether or not schools are functional in reducing inequalities across different geographical areas, students within the same school and attending the same classes receive similar inputs, as they are exposed to the same teachers and curricula. In contexts such as the Italian educational system, students attend the same class with the same group of peers for several school years, which makes even more relevant the equalizer argument. Students learn from their peers and the environment in which they are embedded \citep{angrist2014perils} and develop their human capital also according to their set of unobserved skills. \\
As a consequence of Covid-19 pandemic, governments have implemented school closures and online teaching as one of the main tools to reduce the spread of cases in areas with higher contagion rates, especially when vaccinations were not available and largely uptaken. While these policies may have brought some benefits depending for instance on the timing of adoption \citep{amodio2022schools,vlachos2021effects,Hsiang2020covid}, these undoubtedly came with a potentially high cost for the generation of students attending school \citep{stantcheva2022inequalities}. A growing number of works studies the short and long run effects on a series of outcomes linked to educational losses. Some authors, such as \citet{psacharopoulos2021covid}, report that the long run human capital losses could translate in a decrease of about 8\% in future earnings of the generation exposed to educational losses and expect that this effect will spill over the national GDP. Others, such as \citet{fuchs2020long}, use a structural-life cycle model and highlights that educational losses will induce a decrease in the college attainment rate of about 4\%, a reduction of lifetime earning of 2.1\% and a drop in permanent consumption of 2.1\%.  \\
The present study contributes to the existing literature identifying the educational costs of school closures, expressed as decrease in test scores, for the whole universe of approximately 1,4 million Italian students attending the 5th, 8th and 13th grade of the school cycle during the 2021/22 school year.\footnote{In Italy, schools were full closed and lessons suspended only in the first week of the pandemic (March 2020). After that, school closures implied online learning for all the students. Therefore, in this article, we refer to school closure as the event of moving teaching to online modality.} Using the results from the estimation of an educational production function under equilibrium, we anticipate that the impact will be larger on students enrolled in higher grades, as their function depends more on peers' interactions and less on parents' inputs. The identification relies on a difference-in-difference model in relative time, where the control group is the closest generation before the Covid-19 pandemic. \\
Our estimates at national level suggests a loss between 1.6-4.1\% and 0.5-2.4\% in Mathematics and Italian test scores, respectively. These results speak and contribute to a small but growing number of studies investigating the educational losses by the mean of test scores. In general, these studies find a sizable reduction in learning of  about 4-5\% for the case of Netherlands \citep{engzell2021learning} and Belgium \citep{maldonado2022effect}. These works identify the direct effect of school closures by comparing the human capital dynamics of the affected generation with those of the previous generations, and mostly define the treatment as an aggregated shock. Departing from these studies, our work finds that the impact of school closures is unequal across three dimensions: the school grade the students' are attending, the family background, and the geographical area. \\
With respect to the existing literature \footnote{See for instance the works surveyed in \citep{storey2021meta} and \citep{donnelly2021learning}}, we offer new evidence on the heterogeneous impact of school closures across the geographical territory and estimate the effect of an additional school day of closures on the students' test score. We do so by gathering together original data on the precise number of days of school closures by grade at municipality level for the case study of Sicily. We merge this information with a set of granular data on Covid-19 cases and census information. Findings from this exercise suggest that a 30 days of school closures imply a loss of educational score of about 1\%. This result, however, hides an high heterogeneity across school level, with a peak for students attending high schools. Also, we show a higher effect for students from more disadvantaged backgrounds. We do so by using parents' education and employment status as proxies of parent's ability to recover the gap due by the lack of front-learning. Our findings are also heterogeneous across family job conditions, suggesting that the impact is higher in magnitude for less advantaged students, with either unemployed or blue collar parents. With this, our work provides new evidence on the unequal costs of school closures and online teaching in Italy. A dose-response function suggests that school closures may have a non linear impact on test scores, with students loosing more days of front-teaching observing a sharper decline on their test scores after some thresholds. \\
The remainder of the article is organized as follows. The first part of Section \ref{background} provides a background on the Italian educational system, while the second part discusses about school closures during the pandemic. Section \ref{Data} introduces the data sources and the identification strategy. Section \ref{empirical} presents the production function and the empirical strategy. The first part of Section \ref{results_national} discusses the core results at national level, while results on the impact of additional school days of closure are presented in the sub-section \ref{results_regional}. Section \ref{hetrob} reports the results from additional heterogeneity and robustness tests, while Section \ref{conclusion} presents our conclusive considerations.



%
%

\section{Background}\label{background}

\subsection{The Italian school system}

In 1859, the approval of the Casati Act (\textit{Legge Casati}) assigned the educational responsibilities to the newborn Italian state, with the goal of increasing literacy and making education compulsory. From that moment onward, the Italian educational system underwent to different reforms, designed to increase the level of school attendance, which remained low for several decades, and to enhance the level of human capital of the new generations. Compulsory education was gradually raised from 12 years, with the Orlando Act (\textit{Legge Orlando}) up to 14 years, with the Gentile Act (\textit{Legge Gentile}). In 1923, children, once completed the first five grades of primary education (\textit{scuola elementare}), were allowed to either access to a program foreseeing the middle-school (\textit{scuola media}) and the high-school (\textit{scuola superiore}), or to follow a program of work training (\textit{accesso al lavoro}), intended to prepare them for technical works. This work training was abolished with another major reform in 1962 and, from that period onward, all the Italian students were offered a single program of general education, composed by five years of primary education and three years of middle schools. From 1962 onward, the governments implemented a set of additional reforms aimed at addressing the gender gap in educational access, reforming the high school programs and setting up new specialized high schools for technical professions (\textit{istituti tecnici e professionali}), and raising the compulsory education up to children aged 16 years (\textit{Berlinguer Act}).  \\
Currently, the national educational system is free and compulsory for children between 6 and 16 years old. The system comprises three different levels: primary education (\textit{scuola primaria}), lower-secondary (\textit{scuola secondaria inferiore}), and upper-secondary (\textit{scuola secondaria superiore}). As almost all western Eu countries, the structure of ISCED 1-2 levels is composed by two cycles, instead than a single one, with a common core curriculum for primary education and lower secondary (middle school) education \citep{baidak2019structure}. Once at high school, children may either follow programs of upper-secondary general education, or specialized professional schools, reflecting the traditional dichotomy among secondary general and vocational education. 

The educational outcomes of the Italian system are highly comparable with the one of the other developed countries. From a quantitative perspective, during 1980-2015, the average education in Italy has increased at a rate of almost 2\% per year, following a path similar to other developed countries, such as France and Germany. According to \citet{barrolee2013}, the average education of the population is above 11 years in 2015, a figure slightly lower than OECD average, with a gap of roughly 2 years of average education with respect to UK, USA or major western European countries. Similarly to these, the percentage of Italian population with completed secondary schooling is about 40\%. The gap, therefore, likely derives from the lower share of individuals completing tertiary education in Italy (see the version 2021 of \citet{barrolee2013}). \\

\begin{figure}[h!]
	\begin{center}
		\includegraphics[width=0.85\textwidth]{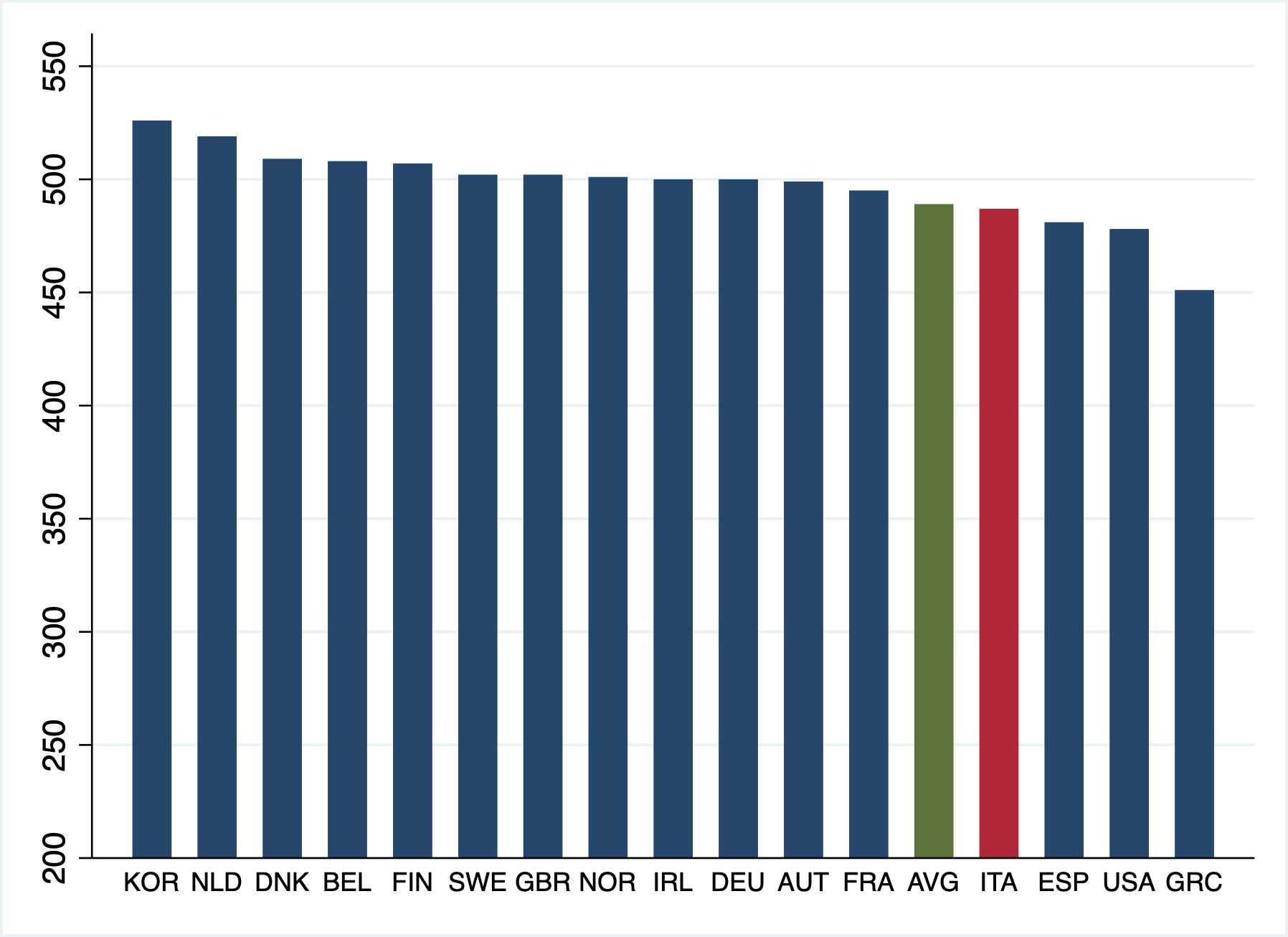}
		\caption{PISA Scores Mathematics - Year 2018}
		\label{fig:PISA}
	\end{center}
\end{figure}

 From a qualitative perspective, the assessment scores in last twenty years show that Italian students hold a level very close to the OECD average, with a score about 6-8\% lower than the frontier, represented by South Korea (Figure \ref{fig:PISA}). 
 Similarly, EUROSTAT suggests that the share of low achieving 15 years old students in Italy is also aligned to the EU average, equal to 23.8\% and 22.9\%, respectively \citet{EUROSTAT2022a}. The Timms scores describe a similar pattern with, for instance, 8th grade Italian students assessing on an average score of 494 in mathematics, a result extremely similar to the average of 500 of OECD countries \citep{fishbein2021timss}. Taken together, this confirms that the Italian educational conditions, especially for the pre-tertiary educational system, may be considered representative of other European and OECD countries.


As for the general socio-economic development, Italy was divided also in terms of education at the time of unification. Illiteracy rates of Southern regions were 4 times higher than those of North-West and three times higher than the country average \citep{bertola2011comparative}, but this gap narrowed down due to the increasing mandatory education. At the present day, some differences persist between the north and south of the country. Data from the national agency for school results evaluation (INVALSI) show that, at the end of high school, students from the south of Italy obtain scores about 20\% lower in Mathematics and Italian with respect to North-West counterparts. Also, in Southern regions the scores have wider variance, about 2-3 times larger than Northern regions, which highlights the wider levels of inequality within the Southern territories. Eurostat data for 2019 show that early school leavers in Sicily are equal to 22.2\% of the total students, about 9 percentage points higher than Italian average, and 12 percentage points higher than the European average \citep{EUROSTAT2022b}.

\subsection{School closures}

%
From March 2020, Covid-19 pandemic has strongly hit the world and a sudden raise in cases was observed in all the western countries. As first response, governments have started to limit the possibility of contagion by enforcing social restrictions (for a complete survey see \citet{Hsiang2020covid}). School physical closures embodied one of the major actions to prevent the diffusion of Covid-19 cases. Schools, indeed, could have represented an occasion of contagion and could have allowed the spread of the virus from asymptomatic children to their household members \citep{amodio2022schools}. 

\begin{figure}[h!]
	\begin{center}
		\includegraphics[width=0.75\textwidth]{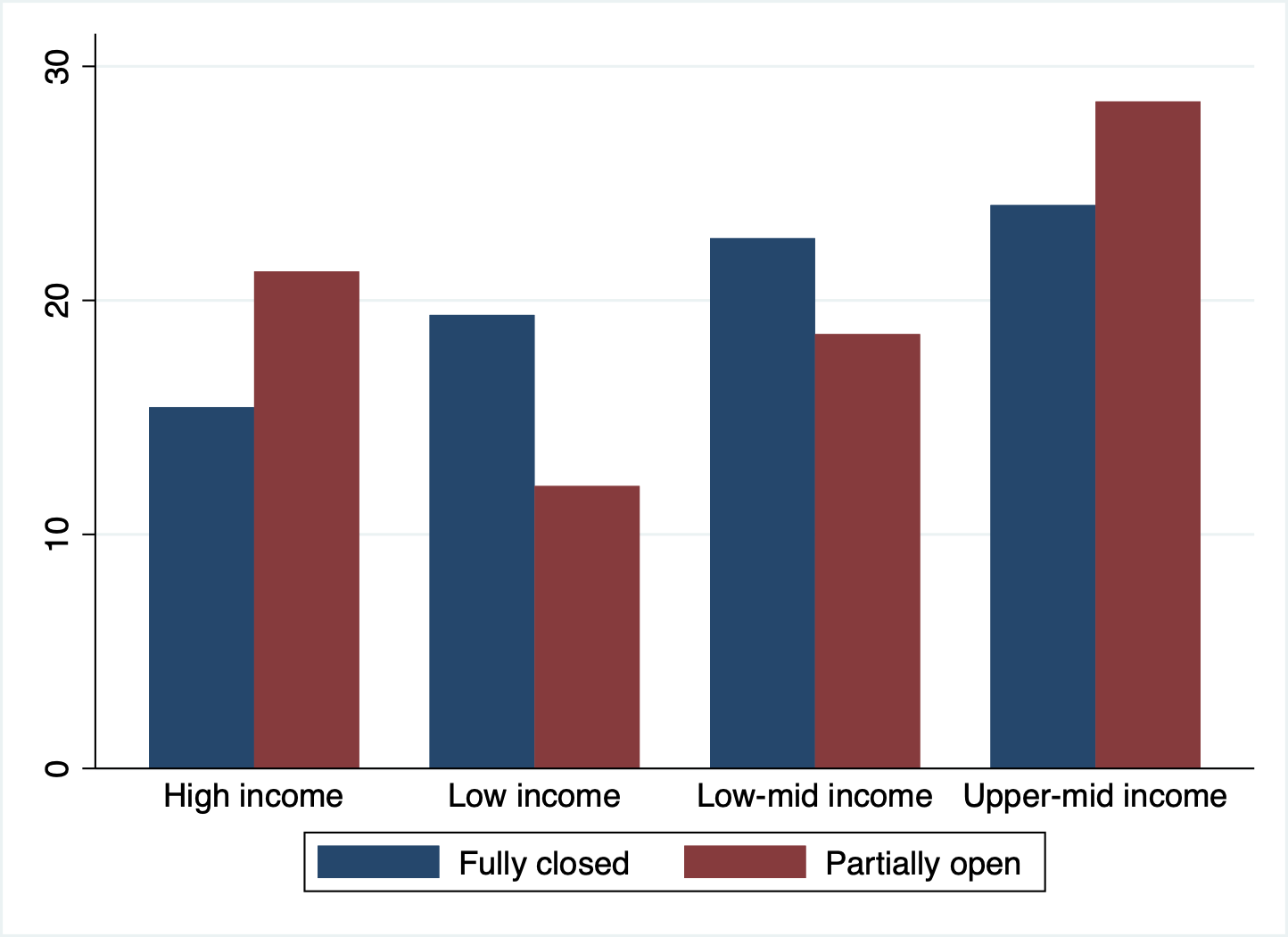}
		\caption{Average number of weeks of school closures from 2020 by country groups. Source: UNESCU (2022).}
		\label{fig:Closures Grades}
	\end{center}
\end{figure}

During the so called first wave of pandemic (March-July 2020), the majority of the developed countries have opted for school closures and have substituted front teaching with the distance learning. However, from the second wave of the pandemic onward, the set of policy answers implemented by the governments depended on the economic and political contexts. As Figure \ref{fig:Closures Grades} shows, the number of weeks without front teaching ranges between 30 and 45. Also the internal composition of fully closed and partially opened schools was very heterogeneous, ranging from high income countries that closing schools for 50\% or more of the weeks during the period, to low income where the percentage of fully closed weeks is one third with respect to high income countries.

Being one of the first countries hit by Covid-19, Italy is among the OECD countries with the highest number of weeks of school closures and distance learning, followed by Greece, Denmark and Finland. This response, however, appears to be driven by internal policy considerations and not by common socio-economic traits. Indeed, even with similar income group countries, it is still possible to observe a substantial heterogeneity in the policies adopted and number of weeks of school closures and distance learning, such as for the case of Switzerland and US where, with a comparable GDP level, we observed completely different approaches.

\section{Data}\label{Data}

The analysis develops on a unique dataset obtained by merging two data sources. The data on test scores derive from the Italian National Institute for the Evaluation of the Educational System (INVALSI). The institute is responsible for evaluating the level of learning of Italian students across all the education-cycle of the Italian educational system every year. In operational terms, the institute conducts on a yearly basis tests based on identical questions on all the universe of the Italian students attending the second, fifth, eighth, tenth, and thirteenth grade of the school.

During the school year 2020/21, about 6,6 millions of Italian students attended primary, middle and secondary schools \citep{ISTAT2022Studenti}. According to the official statistics, the data on test scores for the 2020/2021 school years cover 93.7\% of the students in the targeted grades. Using these data allows, therefore, to estimate the impact on the whole population of students attending the targeted classes, and conduct a set of heterogeneity analyses for different geographical and socio-economic extraction of individuals in this population. The results from the tests are harmonized by the institute using the Rasch model, which is an approach able to weight and model simultaneously the level of difficulty of the questionnaire and the skills of the respondents. The Rasch model attaches more weight, and thus higher scores, to a difficult question answered correctly than to an easy one. This permits to reconstruct more properly the level of heterogeneity in learning across students, especially when compared to alternative simpler approaches, such as counting the number of correct answers. The tests focus on three subjects of study: Italian, Mathematics and English (listening, reading, speaking). The current analysis is based only on the results from the tests on the first two subjects that are administered to all the grades, while the students are tested on their English knowledge only from the 8th grade.\footnote{Unfortunately, English scores are not available for the period preceding the pandemic for the selected grades, so we are unable to test the impact of school closures on these outcomes using panel data}. The INVALSI tests have been conducted every year since 2009/10. The only notable exception is the 2019/20 school year, as the unexpected events from the pandemic impeded to conduct the test. The results from the anonymized test are provided at individual level, and include an identifier for the school, municipality and province. More importantly, each student is associated to a unique panel identifier which allows to follow him/her across all the different grades and to reconstruct his/her test scores during the school cycle. 

\subsection{Building a counterfactual}\label{counterfactual}

Covid-19 and related closures have simultaneously affected an entire generation of student population.  In second part of the 2019/2020 school year, the schools were closed and teaching suspended all over the Italian territory for about one week. After that, in the event of school closure, teaching was moved online for all the involved students. Therefore, in this article, we refer to school closure as the event of moving teaching to online modality. From the second Covid-19 wave onward, school closures occurred with some degree of heterogeneity due to the amount of in-person classes that were moved online, depending on the local trends in Covid-19 cases.  Studying the impact of these on test scores of students is therefore not an easy task due to the absence of a true counterfactual. To tackle this challenge, we compare two cohorts as closest as possible in time, differing only on the experience of school closure and on-line learning. The cohort of treated is the one experiencing the Covid-19-related closures, that is the one observed in the school year 2020/2021. The second cohort is the one observed in 2018/2019, the year before the pandemic has occurred. For both these cohorts, we build a panel of two waves in relative time, adding backwards the same individuals test score results from the closest year available. This is equivalent to take a relative time, where $t=0$ is the time before treatment and $t=1$ the time of the treatment.  The final sample, therefore, is a panel containing the entire universe of Italian students that at time t=1 was attending the 5th, 8th, and 13th grade, and their closet observation in time for $t=0$ .\footnote{We exclude the 10th grade as INVALSI has not administered the test to students of this grade during 2021.}  This means, for instance, that in our panel, all the cohort of students attending the 13th grade in 2020/2021 will also be observed during the 8th grade in 2015/2016. Similarly, we observe the 13th grade control group during the 13th grade in 2018/19 ($t=1$) and the preceding score from the test conducted when this cohort was attending the 8th grade during the school year 2013/14 ($t=0$). Table \ref{tab:tab_school} sum up the panel data collected for each grade of school by treatment status and reports the school-level at the time of the treatment:

\begin{table}[H]
\caption{Treated and control students by grade and year} \label{tab:tab_school}
\begin{center}
\resizebox{0.95 \columnwidth}{!}{%
\begin{tabular}{lccc}
\hline
 \textbf{School level} & \textbf{Grade at t=0 (school year)} & \textbf{Grade at t=1 (school year)}      & \textbf{Treatment status}     \\ \hline
 \multirow{2}{*}{\textbf{High School}} & 8th (2015/16) & 13th (2020/21) &  Treated  group     \\ 
  &  8th (2013/14) & 13th (2018/2019) & Control group \\ \hline

 \multirow{2}{*}{\textbf{Middle School}} & 5th (2017/18) & 8th (2020/21) &  Treated group      \\ 
 &  5th (2015/16) & 8th (2018/2019) & Control group \\ \hline

 \multirow{2}{*}{\textbf{Low School}} & 2nd (2017/18) & 5th (2020/21) &  Treated  group        \\ 
 &  2nd (2015/16) & 5th (2018/2019) & Control group \\ \hline

\end{tabular}
}
\end{center}
\end{table}

\begin{figure}[H]
	\centering
	\minipage{0.5\textwidth} 
	\includegraphics[width=1\linewidth]{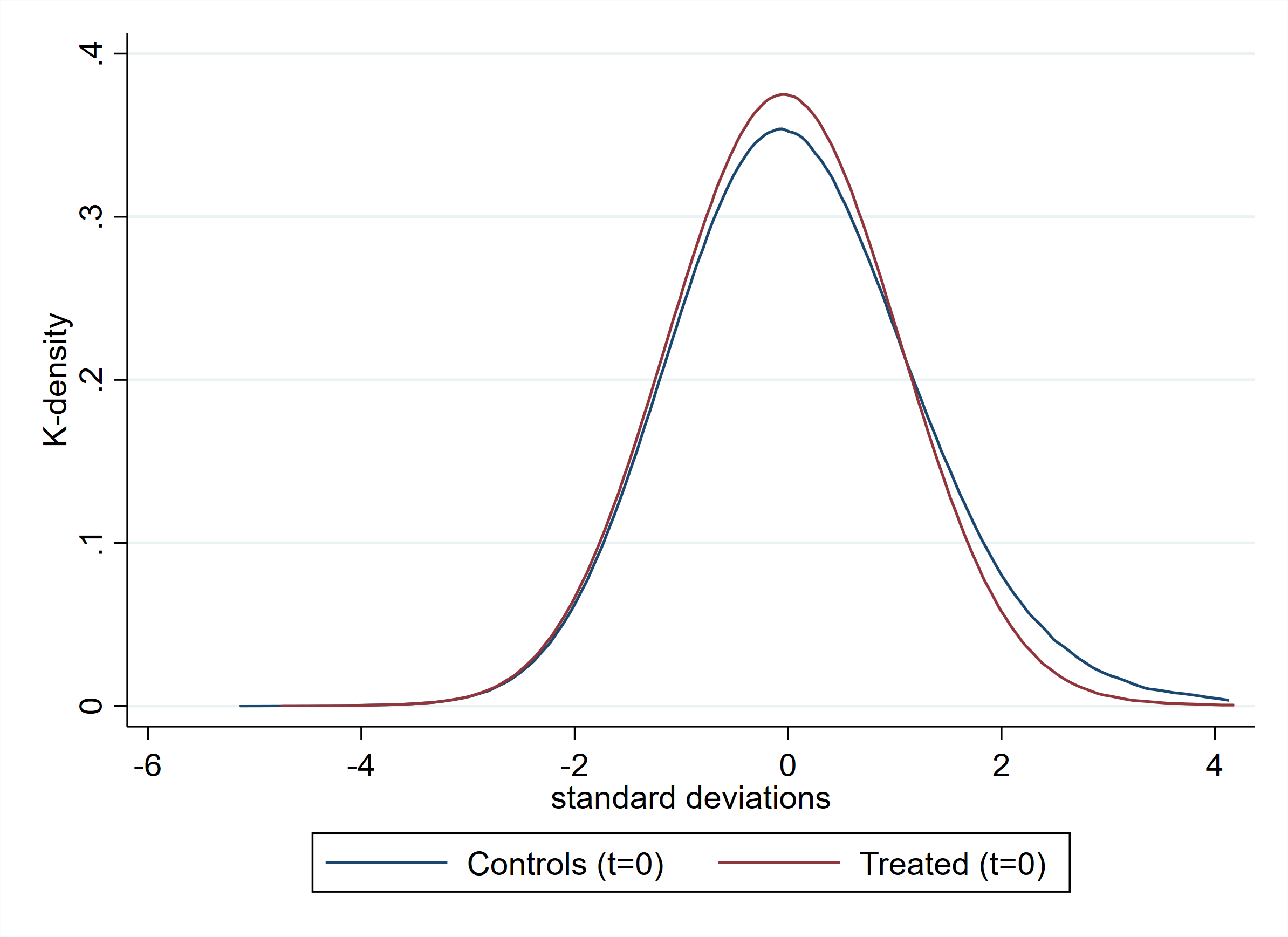}
	\endminipage\hfill
	\minipage{0.5\textwidth}
	\includegraphics[width=1\linewidth]{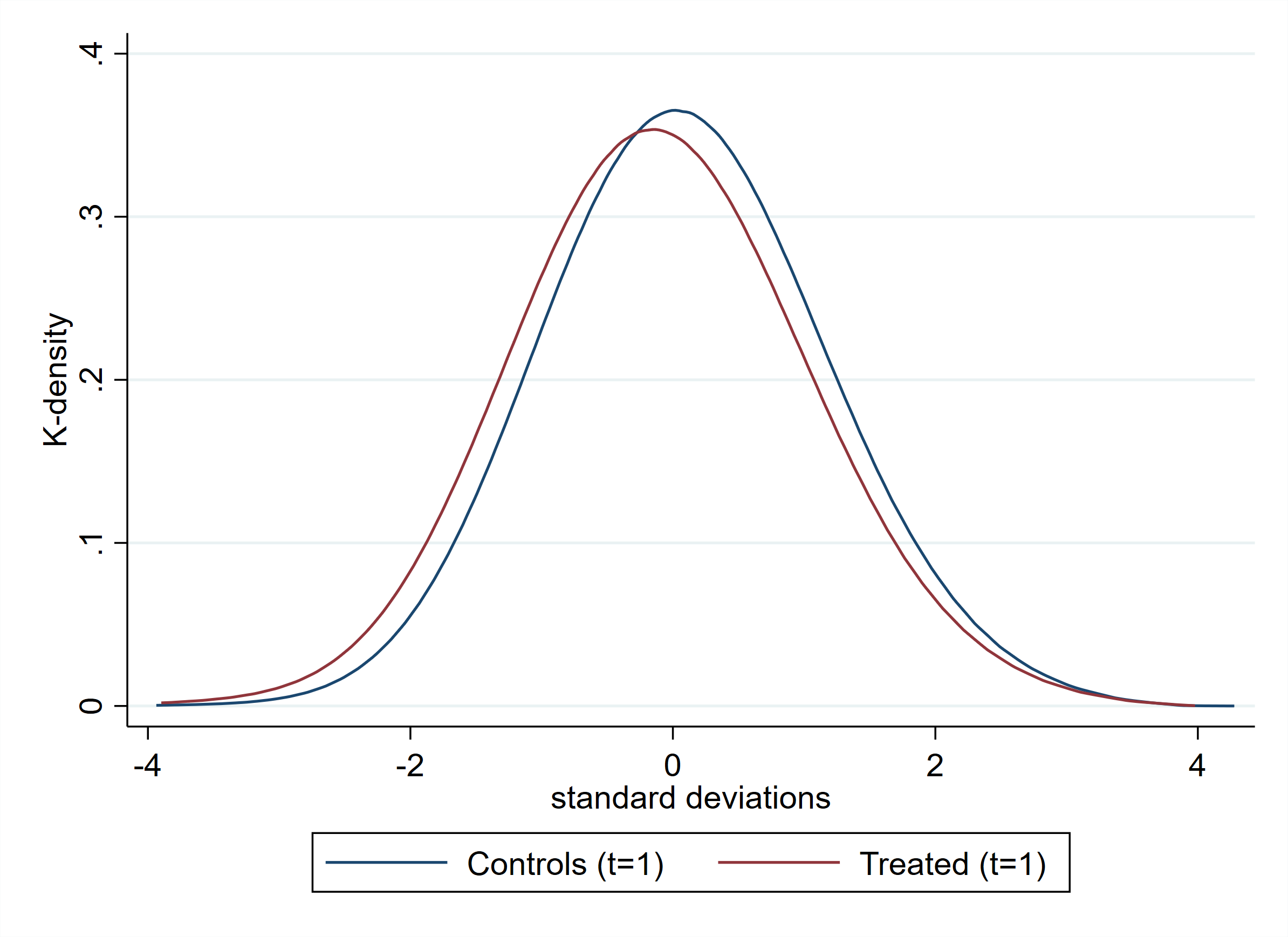}
	\endminipage\hfill
	\caption{Distribution of standardized Mathematic score before ($t=0$) and after ($t=1$) treatment}. 
	\label{Fig_distribution}
\end{figure}

Using these data, we generate a treatment dummy taking value equal to one for individuals in the treatment groups when the relative time is $t=1$. This dummy takes value zero for the individuals belonging to the control group and for those in the treatment group at time $t=0$. Beside the student's score on the test, the INVALSI data provide information on the educational background/title of the parents, their place of birth, being that Italy, European Union, or extra-EU, the parent's employment status and category of employment. In principle, these dimensions are time-invariant and are absorbed by the presence of individual level fixed-effects in the main specification. However we use these to build indicators to be used as controls in a preliminary cross sectional analysis, and interacted with the treatment dummy in the study of the heterogeneous impact of Covid-19 on human capital accumulation. All the continuous variables are standardized by grade and with respect to the same generation distribution at $t=0$, following an approach similar to \citet{abdulkadirouglu2014elite} and \citet{angrist2016stand}. Figure \ref{Fig_distribution} displays the distribution of the Mathematic score before and after treatment, by treatment status and for all the grades in the dataset. As the figure suggests, the treated and control generations are characterized by highly comparable distribution in test scores before treatment ($t=0$). This changes substantially after the treatment occurs ($t=1$), as the distribution of treated generation shows a lower mean and appears more right-skewed, compared to the controls' distribution.

\subsection{Other data sources for the case study in Sicily}\label{counterfactual2}

The variables presented above are sufficient to study the average impact of the Covid-19-related school closures on the learning ability of the students on the 2020/21 cohort. However, calculating only the average generational impact poses the risk of smoothing down the real effect and overlook the variation of this impact, as schools were closed for different time across the Italian territory. Indeed, while during the first Covid-19 wave, schools were all kept closed (March-June 2019/2020) due to central government decision, from the second wave onward schools closures occurred through two mechanisms. A first one was under the control of the school administration, which could decide to move to distance learning a given class if the Covid-19 cases within the same were above three. Second, school closures could have been implemented through a decision of the municipality administration when the number of local cases of Covid-19 were growing above 150 cases per 100,000 inhabitants. This was done by promulgating an administrative act, which was enforced together with other restrictions foreseen from a regulation named \textit{zona rossa} (red-zone). This municipality level act allowed to implement differentiated school closures depending on school grades, for example by keeping in person teaching for low-level schools, while moving high school classes to online teaching. All this suggests a high level of heterogeneity in closures both across municipalities and across school levels within the same municipality. This heterogeneity is likely hidden behind the average impact of our treatment dummy. For this reason, it is necessary to add the precise information on the number of days in which learning was moved online for all the schools in the sample. To the best of our knowledge, it does not exist an harmonized dataset on day of school losses/online for all the universe of schools in Italy. We tackle this challenge by collecting a unique set of information on school closures at grade and school level for the Sicilian territory. This information is coded using the decrees available from the website of the Health Department of the Sicilian Regional Government, which gathers all the acts on school closures for different grade, time, and municipalities\footnote{The decrees are available at the following website: \href{https://www.regione.sicilia.it/la-regione-informa/covid-19-ordinanze-disposizioni-attuative}{https://www.regione.sicilia.it/la-regione-informa/covid-19-ordinanze-disposizioni-attuative}}. For the cross-sectional analysis, we integrate this information with records on population level obtained from the national census data, conducted by the national statistical office (ISTAT) in 2011. In a robustness test, we check whether our results are driven by reverse causalities or spurious correlation by instrumenting the number of days of closure, by using two variables describing the long-term Covid-19 pattern at municipality level. These instruments are the number of Covid-19 cumulative cases for a given municipality and on the variance of cases across time in a given municipality. As for the other continuous variables, also the indicator on day of school closures is standardized around zero using its mean and standard deviation. 
 Table \ref{tab:tab_school2} reports the statistics of the variables for the national sample, while Table \ref{tab:tab_school3} displays the statistics for the Sicilian sample. As these tables show, the average difference between treated' and controls' test scores is similar for both national and Sicilian data, around 7 points for Math and 4 for Italian, even with different starting levels. Also, following the design of this exercise, at relative time $t=0$ the sample does not contain any student attending the 13th grade, as all the grades are lagged with respect to time $t=1$. 

\begin{table}[H]
	\caption{Summary statistics for the national sample by relative time} 
	\label{tab:tab_school2}
	\begin{center}
\resizebox{1.03 \columnwidth}{!}{%
		\begin{tabular}{lcccccc|cccccc}
		\hline
		 &  \multicolumn{6}{c}{\textbf{Relative time=0}}  & \multicolumn{6}{c}{\textbf{Relative time=1}}  \\ \hline
		 & \multicolumn{2}{c}{\textbf{Full Sample}}  &  \multicolumn{2}{c}{\textbf{Treated}}  &  \multicolumn{2}{c}{\textbf{Control}} &  \multicolumn{2}{c}{\textbf{Full Sample}}  &  \multicolumn{2}{c}{\textbf{Treated}}  &  \multicolumn{2}{c}{\textbf{Control}} \\ 
      Variables & Mean & SD & Mean & SD & Mean & SD & Mean & SD & Mean & SD & Mean & SD \\ \hline
Score in Math test & 211.14 & 40.87 & 209.87 & 37.89 & 212.17 & 43.10 & 202.18 & 38.84 & 198.22 & 39.71 & 205.97 & 37.61 \\ 
Score in Italian test & 208.42 & 39.81 & 206.65 & 37.72 & 209.85 & 41.35 & 202.96 & 37.66 & 200.67 & 37.93 & 205.15 & 37.27 \\ 
Peers' score in Math test & 210.96 & 24.67 & 208.16 & 20.23 & 213.22 & 27.52 & 202.01 & 23.97 & 198.21 & 24.55 & 205.64 & 22.83 \\ 
Peers' score in Italian test & 208.21 & 21.33 & 204.84 & 18.01 & 210.92 & 23.31 & 202.78 & 21.72 & 200.66 & 22.76 & 204.80 & 20.47 \\ 
Parents' year of education & 10.78 & 6.02 & 11.49 & 5.73 & 10.21 & 6.18 & 10.90 & 6.01 & 11.46 & 5.74 & 10.37 & 6.20 \\ 
Student repeating the year (1=yes) & 0.02 & 0.14 & 0.02 & 0.13 & 0.02 & 0.14 & 0.02 & 0.13 & 0.02 & 0.13 & 0.02 & 0.13 \\ 
5th grade - Low School (1=yes) & 0.36 & 0.48 & 0.39 & 0.49 & 0.34 & 0.47 & 0.33 & 0.47 & 0.32 & 0.47 & 0.35 & 0.48 \\ 
8th grade - Middle School  (1=yes) & 0.32 & 0.47 & 0.28 & 0.45 & 0.36 & 0.48 & 0.39 & 0.49 & 0.40 & 0.49 & 0.37 & 0.48 \\ 
13th grade - High School  (1=yes) & 0.00 & 0.00 & 0.00 & 0.00 & 0.00 & 0.00 & 0.28 & 0.45 & 0.28 & 0.45 & 0.28 & 0.45 \\ \hline
Observations & \multicolumn{2}{c}{2,306,857} & \multicolumn{2}{c}{1,127,790} &\multicolumn{2}{c}{1,179,067} &  \multicolumn{2}{c}{2,306,857}  & \multicolumn{2}{c}{1,127,790} &\multicolumn{2}{c}{1,179,067}  \\ \hline
\end{tabular}
}
\end{center}
\end{table}

\begin{table}[H]
	\caption{Summary statistics for the Sicily sample by relative time} 
	\label{tab:tab_school3}
	\begin{center}
		\resizebox{1.03 \columnwidth}{!}{%
			\begin{tabular}{lcccccc|cccccc}
				\hline
				&  \multicolumn{6}{c}{\textbf{Relative time=0}}  & \multicolumn{6}{c}{\textbf{Relative time=1}}  \\ \hline
				& \multicolumn{2}{c}{\textbf{Full Sample}}  &  \multicolumn{2}{c}{\textbf{Treated}}  &  \multicolumn{2}{c}{\textbf{Control}} &  \multicolumn{2}{c}{\textbf{Full Sample}}  &  \multicolumn{2}{c}{\textbf{Treated}}  &  \multicolumn{2}{c}{\textbf{Control}} \\ 
				      Variables & Mean & SD & Mean & SD & Mean & SD & Mean & SD & Mean & SD & Mean & SD \\ \hline
				Score in Math test & 216.40 & 42.81 & 211.76 & 37.36 & 221.33 & 47.44 & 189.39 & 38.75 & 185.46 & 40.47 & 193.57 & 36.38 \\ 
				Score in Italian test & 209.93 & 42.25 & 204.86 & 38.44 & 215.33 & 45.33 & 192.42 & 38.46 & 190.48 & 39.20 & 194.49 & 37.55 \\ 
				Days of school closure & 0.00 & 0.00 & 0.00 & 0.00 & 0.00 & 0.00 & 72.85 & 77.30 & 141.38 & 43.67 & 0.00 & 0.00 \\ 
				Peers' score in Math test & 221.66 & 35.24 & 217.13 & 28.79 & 226.49 & 40.44 & 193.17 & 30.38 & 189.11 & 31.72 & 197.48 & 28.25 \\ 
				Peers' score in Italian test & 216.59 & 32.22 & 212.50 & 28.40 & 220.94 & 35.32 & 196.55 & 29.36 & 195.44 & 31.97 & 197.73 & 26.26 \\ 
				Parents' year of education  & 9.69 & 5.99 & 9.99 & 5.93 & 9.37 & 6.04 & 9.69 & 5.99 & 9.99 & 5.93 & 9.37 & 6.04 \\
				Student repeating the year (1=yes) & 0.01 & 0.08 & 0.01 & 0.09 & 0.00 & 0.06 & 0.01 & 0.12 & 0.01 & 0.12 & 0.02 & 0.12 \\
				Parents are unemployed or blue-collars (1=yes) & 0.19 & 0.39 & 0.20 & 0.40 & 0.19 & 0.39 & 0.19 & 0.39 & 0.20 & 0.40 & 0.19 & 0.39 \\ \hline
				Foreigner parents & 0.03 & 0.18 & 0.04 & 0.19 & 0.03 & 0.16 & 0.03 & 0.18 & 0.04 & 0.19 & 0.03 & 0.16 \\ 
				Class size & 19.30 & 2.93 & 19.37 & 2.91 & 19.23 & 2.95 & 20.27 & 2.84 & 20.36 & 2.81 & 20.17 & 2.86 \\ 
				Municipality's population & 260,887 & 41,5004 & 257,873 & 410,850 & 264,091 & 419,352 & 285,908 & 425,487 & 282,295 & 421,231 & 289,747 & 429,933 \\ 
				Number of classes in the school & 13.99 & 7.36 & 14.18 & 7.44 & 13.80 & 7.27 & 21.56 & 16.15 & 21.82 & 16.14 & 21.28 & 16.17 \\
				Covid-19 cases per inhabitant & 0.00 & 0.00 & 0.00 & 0.00 & 0.00 & 0.00 & 0.02 & 0.03 & 0.04 & 0.03 & 0.00 & 0.00 \\
				Municipality's variance of Covid-19 cases over time & 0.00 & 0.00 & 0.00 & 0.00 & 0.00 & 0.00 & 20,270 & 25,952 & 39,340 & 23599 & 0.00 & 0.00 \\  \hline
				Observations & \multicolumn{2}{c}{167,183} &  \multicolumn{2}{c}{86,140} & \multicolumn{2}{c}{167,183} &  \multicolumn{2}{c}{86,140} & \multicolumn{2}{c}{167,183} &  \multicolumn{2}{c}{86,140}  \\ \hline
			\end{tabular}
		}
	\end{center}
\end{table}

For the school year 2020/21, the INVALSI data include information on so 1,9 million students\footnote{The entire population of students in the selected grades is  2,105,000, but only 93.7\% of students attended the test.}. Excluding the 2nd grade due to the lack of counterfactual reduces our sample to 1,379,000 students. Among them, we consider students that completed tests in both subjects (Math and Italian), so that the final national sample reduces to about 1,2 million of students.

\section{Production function and empirical strategy}\label{empirical}

\subsection{Educational production function}\label{Strategy}
The empirical approach of this analysis grounds on a standard education production function, on the basis of \citet{hanushek1970production} and \citet{hanushek1979conceptual}, which can be formalized as follows:

\begin{equation}
Y_{i,t}=f\left(B_{i,t}, S_{i,t}, \bar{P}_{i,t}, I_{i}\right )	
\end{equation}

Where $Y_{i,t}$ is a measure of achievement for student \textit{i} at time \textit{t}, \textit{B} are family background inputs, \textit{S} is the school input received by the student at time (year) \textit{t}, $\bar{P}$ is the peers' influence on the student's educational attainment, and \textit{I} are the time invariant, idiosyncratic abilities. Note that, as in \citet{bowles1970towards}, the term \textit{Y} may represent both the level of educational attainment or its value added function, i.e. the difference between the first and last period of measurement. \citet{polachek1978educational} also clarify that the functional form of equation 1 depends on the underlying assumptions linked to the inputs. For example, selecting a Cobb Douglas would imply to assume either complementarity or substitutability between the inputs. Assuming a simple linear relationship between these inputs and the dependent variable, we can rewrite this relationship as a simple OLS specification taking the following form:
 
 \begin{equation} 
 \label{equ1}
Y_{i,g,s}=\alpha + \beta \Pi_{i}  +\lambda \bar{P}_{i,g,s} +S_{s}+\epsilon_{i}  
\end{equation}
 
 Where the dependent variable is the educational score of student \textit{i} attending the grade \textit{g} in school \textit{s}. This variable is regressed on $\Pi_{i}$, denoting parent's years of education, on the input of the classmates/peers $\bar{P}_{i,g,s}$, measured as leave out mean of the class-mates score and thus varying at individual level, and on the school input $S_{s}$, captured by a set of school-dummies. Finally, $\epsilon_{i}$ is the error term, clustered at an individual-level. \\
 The above equation provides information on the relationship between these inputs and the educational outcome when the system is under equilibrium conditions. We refer to equilibrium conditions as the ones that were in place before Covid-19 pandemic and their related school-closures took place. When run by grade, the results of this statistical association may inform about the unequal impact of the school-closures. For example, if the educational score of pupils from the 5th grade will be more dependent on the school background rather than on the peers, we will expect that online teaching will have had a limited effect on their educational score. This is the case for the students attending 13th degree, who are on average 18-19 years old, they likely are more self-sufficient in their homework and more likely to do these with their classmates/peers.\footnote{This derives by the fact that the educational system in Italy foreseen the same set of courses for the same group of students. Differently from US, where students can select their courses, Italian students are enrolled under a class and will share the same subjects with the same group of student for the whole school-level (low-school, mid-school, high-school).}. We run this cross-sectional regression using the 2019 cohort to understand, therefore, the potential level of vulnerability of students to school-closures depending on their grade.\footnote{We acknowledge that unobserved time-invariant characteristics may be contributing to this association. However, given that the educational background of the parents is mostly time-invariant, it is not possible to run a OLS-FE specification using both waves of the 2019 data. We have thus decided to rely on a simple OLS model, but we also run the estimates from a random-effect model with Mundlak correction, which remains consistent with the results presented in the main part of the analysis (available upon request).}
 
 

\begin{figure}[H]
	\centering
	\minipage{0.5\textwidth} 
	\includegraphics[width=1\linewidth]{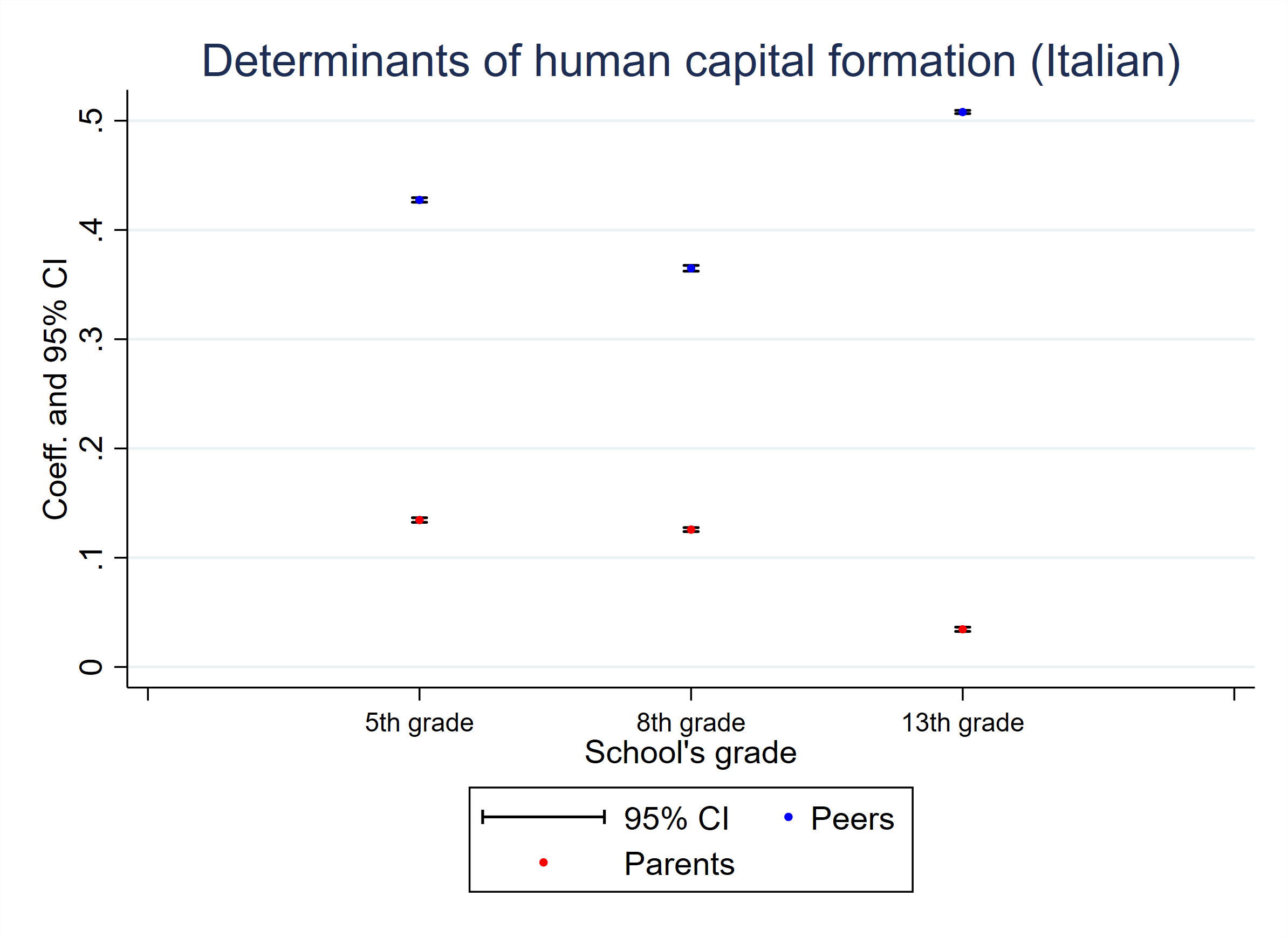}
	\endminipage\hfill
	\minipage{0.5\textwidth}
	\includegraphics[width=1\linewidth]{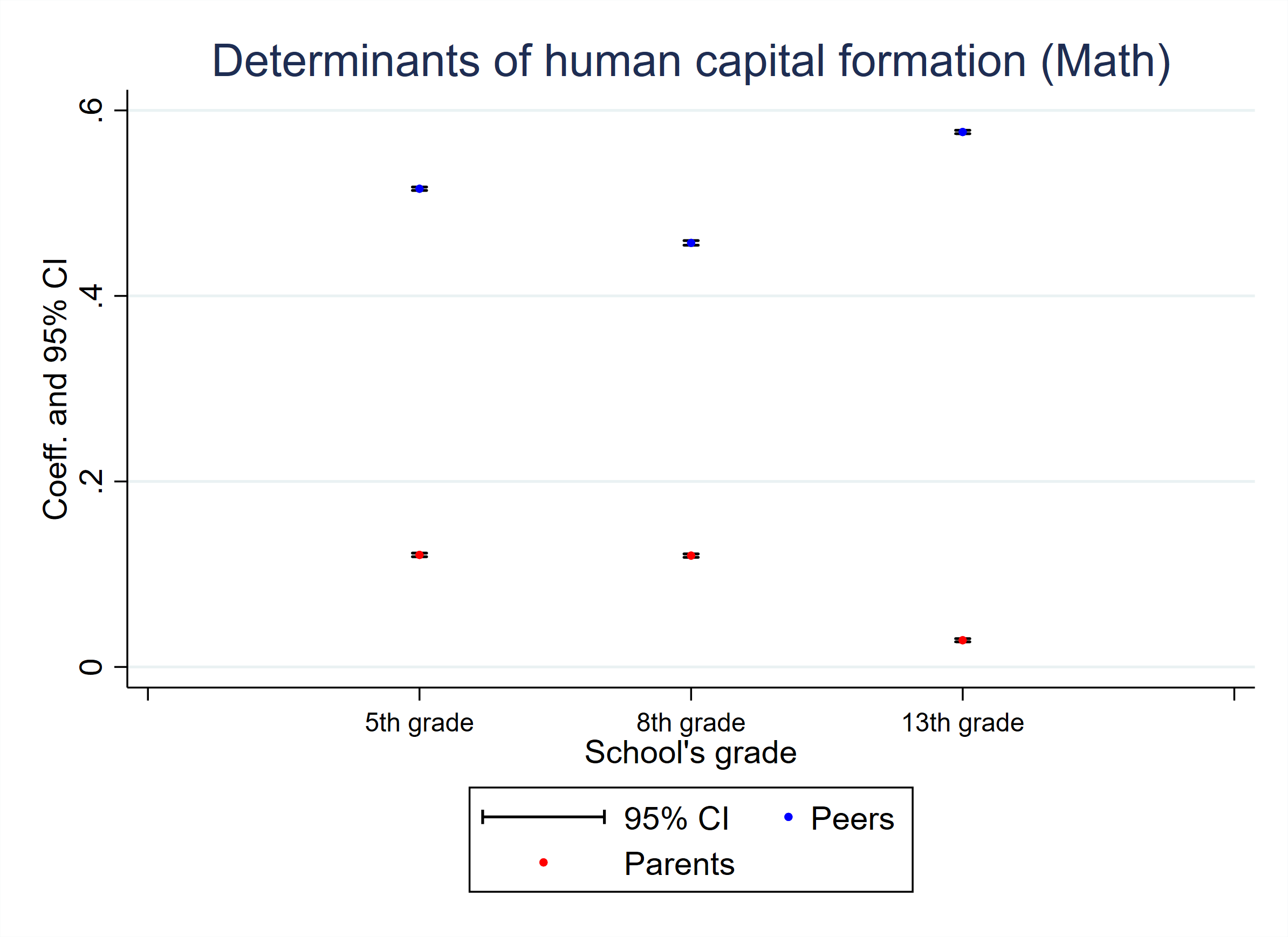}
	\endminipage\hfill
	\caption{Coefficients of education production inputs by standardized variables: Italian and Math}. 
	\label{Fig_coefficients}
\end{figure}

Figure \ref{Fig_coefficients} below reports the parents and peers' coefficients when running the specification \ref{equ1} across grades for Italian and Mathematics scores. In this specification, the continuous variables are transformed using natural log to provide a visual intuition about the relative weight of different coefficients, but the results are extremely similar with standardized variables. The magnitude of parents' background coefficient is similar for the 5th and 8th grade, while it decreases substantially for the 13th grade. In contrast, peers' coefficient appears to increase for students attending the 13th grade. Taken together, these coefficients suggest that, in a situation where parents should compensate the lack of in front teaching, this effect is likely to be higher for students attending the lower grades of schooling. Also, since students attending higher grades are more likely to benefit from peers' inputs, moving classes to online learning has the potential to weaken even more these students.

\subsection{Closures effects on educational scores: a diff-in-diff approach}
 
 We test the average impact of school closures on educational outcome using a panel setting and comparing the cohort of students before the pandemic as control group and apply a difference-in-difference approach in relative time. We consider two cohorts $j=\{p, c\}$ observed at $t=0,1$. The cohort $p$ correspond the the population of treated students which were affected by the school closures. For these students, $t=1$ corresponds to the school year $2020/21$, while $t=0$ depends on the school grade, as reported in Table \ref{tab:tab_school}. Similarly, with cohort $c$ we refer to the population of control students observed during the school year 2018/19 ($t=1$), and the respective preceding grade available from the INVALSI test (see Table \ref{tab:tab_school}). Following \citet{engzell2021learning}, and considering that parents' education is time-invariant and is absorbed by the individual fixed-effects, we formalize the approach as follows:
 
 \begin{equation}
 \label{eq2}
Y_{i,j,g,t}= \beta_{0}+ \delta T_{i,t} +\lambda \bar{P}_{i,j,g,t}+\Phi_{i}+\Theta_{t}+\Gamma_{g} +\varepsilon_{i,t}	
\end{equation}

Where the educational score of student $i$ of the cohort $j$ attending the school grade $g$ at time $t$,  $\beta_{0}$ denotes the intercept, and dummy $T_{i,t}$ is the treatment, taking value equal to 1 for students belonging to the cohort $p$ at time $t=1$ and zero otherwise. Also, the set of controls include classmates score $\bar{P}_{i,j,g,t}$, which is individual specific as it is calculated as a leave-out average of the class score, a set of student-level fixed effects $\Phi_{i}$, two relative time dummies $\Theta_{t}$ and a set of grade-level dummies $\Gamma_{g}$, while $\varepsilon_{i,t}$ denotes the error term clustered at individual level.

We also investigate the specific impact of additional school closure days on the students' Mathematics and Italian scores. To do so, we slightly modify the equation \ref{eq2}, substituting the treatment dummy with a continuous indicator, as follows:

\begin{equation}
	\label{eq3}
	Y_{i,m,j,g,t}= \beta_{0}+ \eta DaysClosures_{m,g,t} +\lambda \bar{P}_{i,j,g,t}+\Phi_{i}+\Theta_{t}+\Gamma_{g} +\upsilon _{i,t}	
\end{equation}

Where the $DaysClosures_{i,t}$ denotes a variable capturing the days of school closures and online learning for a given grade $g$  attended by the student $i$ in municipality $m$, and $\eta$ being the associated coefficient. As for the other continuous variables, also this indicator is standardized to have zero mean and unitary variance. The other components of the model remain identical and, as before, the specification includes individual fixed effects and time dummies.
Finally, to investigate the potential heterogeneity behind the average impact, we extend the equation \ref{eq3} by including the interaction terms between the days of school closures and the peers' score, and between the days of school closures and parents' year of education. This further specification takes the following form:

\begin{equation}
\begin{aligned}
	Y_{i,m,j,g,t} = & \beta_{0}+ \eta DaysClosures_{m,g,t} + \Pi_{i} * \zeta DaysClosures_{m,g,t} \\
	& +\bar{P}_{i,g,s} * \iota DaysClosures_{m,g,t}
	 +\lambda \bar{P}_{i,j,g,t}+\Phi_{i}+\Theta_{t}+\Gamma_{g} +\upsilon _{i,t}	
\end{aligned}
\end{equation}
 
Where the $\zeta$ and  $\iota$ are the two coefficients of interest linked to the interactions between the days of school closures and the peers' score, and between the days of school closures and parents' year of education, respectively.

\section{Effect of school closures on educational scores} \label{results_national}
\subsection{Average impact at a national level} \label{results_national1}
Table \ref{tab:result1} introduces the results of the Mathematics and Italian score specifications by incrementally account for cross-sectional and panel differences. For the sake of comparison, column 1 and 2 report the results from a treatment effect consisting in an OLS specification for sample observed only at relative time $t=1$, including relative time and school dummies. Column 3, instead, reports the results from the baseline specification in equation \ref{eq2}. The first two columns differ on the time of observation of the peers' score: while the first one uses a lagged variable, the second one employs a contemporaneous peers' score indicator. This is done to test whether controlling for the contemporaneous indicator substantially affect the impact of school closures, given that school closures affected all the peers at time $t=1$ \footnote{Unfortunately, since we do not have more than two waves per student, it is impossible to test a DiD specification including the lagged value of the peers' score.}. As the first two columns suggest, the impact of school closures is negative and significant in both cases, but part of school closure effect is absorbed by the contemporaneous peers' score indicator. A similar result emerges when running our baseline specification including individual fixed effects and relative time dummies, reported in column 3 and 6, respectively. Overall, the results from suggest a negative and significant impact of school closures on the educational scores of the affected students. Being all continuous variables standardized, one should account for the value of the original standard deviation of the test score, which is equal to about 38.85 for the case of Mathematics score, and the average of Mathematics score, equal to 202.18 in the cross section.


Taken together, and assuming a yearly linear production function, coefficients of -0.215 and -0.085, with a standard deviation of implies a loss in Mathematics score between 1.8 and 4.1\% \footnote{This is calculated multiplying the coefficient for the standard deviation and dividing the result with the average of reference.}. For the DiD specification, the magnitude of the loss is about 1.6\%. taking into account these differences in the specifications, our model suggests that the average impact of school closures at national level is bounded between 1.6-4.1\%. This finding is in line with what found in other contexts. For example, for the case of Netherlands, \citet{engzell2021learning} find an impact of 0.08 standard deviations, corresponding to about 3\% in educational loss. A similar result holds for the Italian score, with a negative and significant impact of school closures on educational scores bounded between 0.5 and 2.4\%.

\begin{table}[H]
	
	\begin{center}
		\resizebox{0.95 \columnwidth}{!}{
			\begin{threeparttable}
				\caption{Effect of Covid-19 closures on educational attainment} \label{tab:result1}
				\begin{tabular}{lcccccc}
					\hline
					& \multicolumn{3}{c}{\textbf{Score in math test}}  & \multicolumn{3}{c}{\textbf{Score in italian test}} \\
					& Cross-section & Cross-section & Panel - Diff. in Diff. 	& Cross-section & Cross-section & Panel - Diff. in Diff. \\
					& (1) & (2) & (3) & (4) & (5) & (6) \\
					\hline
					&  & &  &  &  \\ 
					
					\textbf{School closures}  & -0.215*** & -0.085*** & -0.047*** & -0.127*** & -0.081*** & -0.025*** \\
					(Treated=1 X t=1)  & (0.001) & (0.001) & (0.001) & (0.001) & (0.001) & (0.001) \\
					\hline
					
					Peers' score in math test (lag)  &  0.058***  &   &  &  &  &  \\
					& (0.000)  &   & &  &  &  \\
					
					Peers' score in math test &  & 0.430*** & 0.449*** &  &  &  \\
					 &  & (0.001) & (0.000) &  &  &  \\

					Peers' score in Italian test (lag) &  &  &  & 0.043*** &  &   \\
					&  &  &  & (0.000)  &  &  \\
					
					Peers' score in Italian test  &  &  &  &  & 0.339*** & 0.359*** \\
					&  &  &  &  & (0.001) & (0.001) \\

					Student repeating the year (1=yes) & -0.401*** & -0.361*** & 0.012 & -0.554*** & -0.517*** & 0.011 \\
				 & (0.005) & (0.004) & (0.013) & (0.005) & (0.005) & (0.014) \\ \hline 
					
					\textbf{Grade dummies} & Yes & Yes & Yes  & Yes & Yes & Yes  \\  
					\textbf{School Dummies} & Yes & Yes & No  & Yes & Yes & No    \\  \hline 
					\textbf{Student FEs} & No & No & Yes & No & No & Yes  \\
					\textbf{Relative-time dummies} & No & No & Yes  & No & No & Yes \\

					\textbf{R-squared}& 0.194 & 0.297 & 0.362 & 0.164 & 0.231 & 0.231  \\
					\textbf{Observations} & 2,306,857 & 2,306,857 &  4,613,714 & 2,306,857 & 2,306,857 & 4,613,714 \\
					\textbf{Number of students} & 2,306,857 &2,306,857 & 2,306,857 & 2,306,857 & 2,306,857 &  2,306,857  \\ \hline
				\end{tabular}
				
				\begin{tablenotes}
					\item \textit{Notes: }{the table reports the estimates from an OLS cross-sectional model (columns 1-2 and 3-4) and Two-way fixed effect model (columns 3 and 6) on the impact of Covid-19 related closures on educational score of students. The main explanatory variable is a dummy activating for the treated group observed during the 2021/22 school year. The OLS-FE specification includes student-level fixed effects and relative time dummies. For more details on the treated and control group see section \ref{Data} and table \ref{tab:tab_school}. Robust standard errors clustered at student-level are reported in parentheses. Level of significances are ***, **, and * for significance at 1\%, 5\%, and 10\%, respectively.}{\footnotesize\par}
				\end{tablenotes}
			\end{threeparttable}
			
		}
	\end{center}
\end{table}

\subsection{Impact of additional days of school closures}\label{results_regional}

While the average national impact aligns with other works, it may still hide a wide heterogeneity across grades, parents' background, peers' human capital, and geography. We do so by taking advantage of a unique dataset on days of school closures by grade and municipality in Sicily. Schools were closed with different degrees depending on the local trend of Covid-19 cases, which varied substantially across the regional territory. Depending on the trend in cases and occupancy of intensive care units, the decisions about closures were suggested by the municipalities and taken by the regional government. This allows to estimate the impact of an additional day of school closures and, therefore, moving from an average treatment specification to a continuous treatment one.

\begin{figure}[h!]
	\begin{center}
		\includegraphics[width=0.80\textwidth]{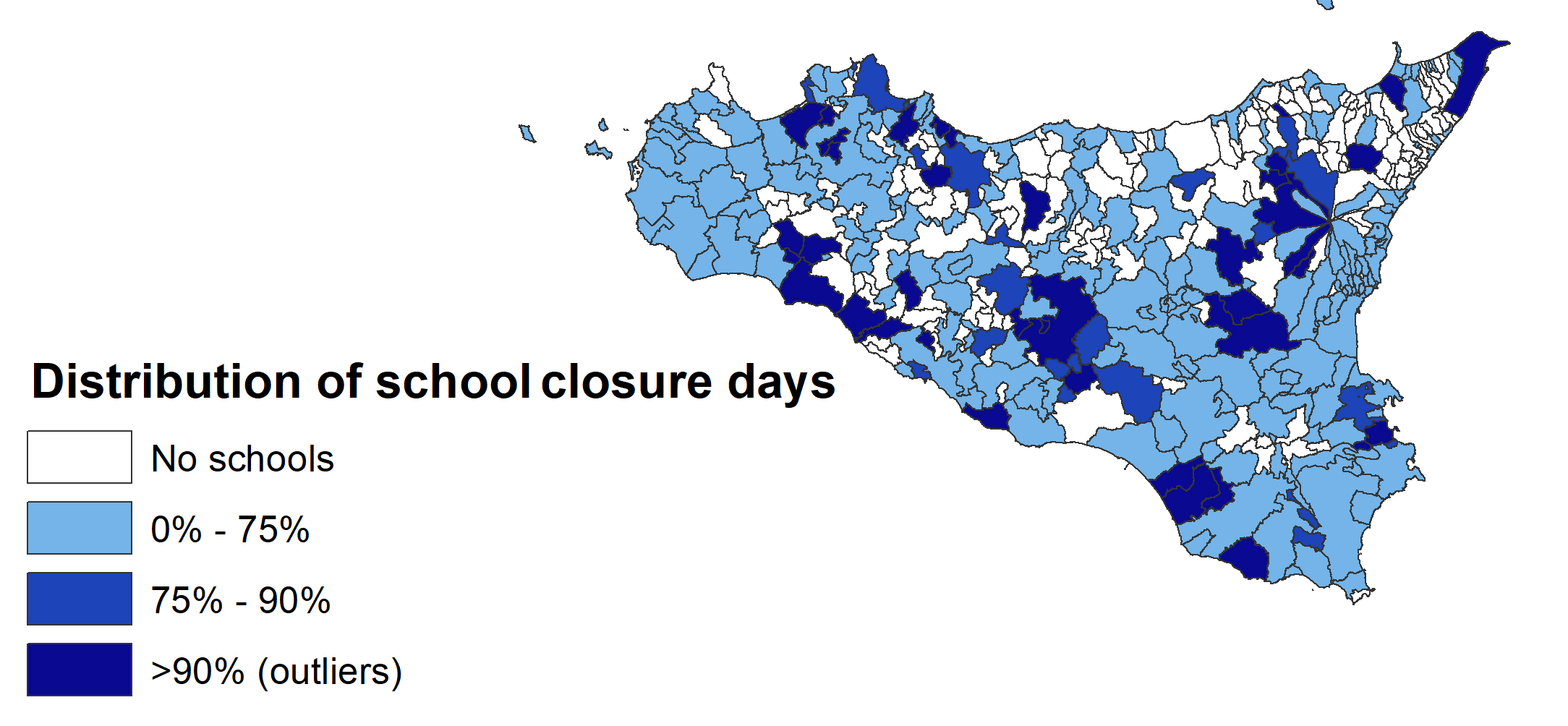}
		\caption{Average days of closures by municipalities}
		\label{fig:Closures Sicily}
	\end{center}
\end{figure}

Figure \ref{fig:Closures Sicily} shows the map of average days of school closure by municipalities.\footnote{The spatial heterogeneity is similar when mapping school closures for separate grades and these maps are available upon request.} White areas represent municipalities without schools, which are usually low populated. These hold an average of 150 inhabitants, a maximum of 482 inhabitants. If we consider the regional population by age, these municipalities should hold on average 21 students and a maximum of 67. Dividing these by the number of grades in the Italian school systems, the municipalities without schools hold maximum 5 individuals in school-age per grade. 

Figure \ref{fig:Closures Sicily} shows that data are skewed on the right tail of the distribution. Indeed, one fourth of municipalities are associated to a number of days of school closures substantially higher than the remaining. Also, the percentage difference between the 90th percentile and the mode value of school closures is about 7-12\%,\footnote{We obtained this share by dividing the days of school closures by a total of 200 days for Italian schools according \citet{baidak2019structure}}. When considering the 95th percentile of school closure days, the difference from the mode is between 14 and 25\%, depending on the grades. This means that applying a binary treatment indicator, as in Section \ref{results_national1}, may hide a large degree of heterogeneity across schools and municipalities.


Table \ref{tab:days_closure} displays the results when replicating the main specification but substituting the binary indicator with the continuous one. As before, the identification in the panel specification relies on a between cohort-comparison similar to what we had at a national level. Results from the baseline specification in column 1 suggest that an increase in 1 standard deviation in schools days is associated to a decrease in the average Mathematics and Italian score. As for the treatment dummy, the magnitude of the coefficient of school closure days increases substantially when accounting for the educational background of the parents and the level of the peers' score. Considering that the standard deviation of the days of school closures and of the Mathematics score are equal to 38.28 and 43.01, respectively, while the average of the same score is 202.9, the results in column 2 suggest that 30 additional days of school closures decreases the score of about 1\%\footnote{This result is obtained calculating the impact of one day of school closures on the standardized score and re-scaling the result using the mean and the standard deviation of the score multiplied by 30.}. A similar result emerges from the specification considering the Italian score, with 30 additional days of closures associated to a decrease in the score of about 0.97\%. To put these results into perspective, the difference in days of closure between the 25th and 90th percentile of our sample is 97 days, which correspond to a decrease of about 3.2\% in the Mathematics score, and 3.1\% in the Italian score.

It is important to stress more the heterogeneous impact that an additional day of school closure may have, depending on the parental and peers' background. Results in column 2 and 3 of Table \ref{tab:days_closure} allows to do so, as they include the interaction terms between parents' year of education, peers' score, and the continuous treatment indicator. When considering this richer model, the effect of school closures appears slightly higher compared to column 1, and remains negative and significant. Also, the coefficients of the interaction terms report the expected sign and magnitude. On one side, the coefficients of interaction between school closures and more educated parents is positive and significant, suggesting that highly educated parents appear to compensate the negative impact of school closures. This remains, however, just a partial compensation, given that the coefficient of the interaction term is lower than the one linked to the indicator on days of closures. On the other side, note that the coefficient of the interaction between peers' with higher educational scores and school closures, is negative and significant, supporting the hypothesis that school closures affected more the students that were sharing the classes with the best performing peers. 

Finally, in column 5 and 6 of Table \ref{tab:days_closure}, we pool the sample across the school level, to study whether school closures have disproportionately impacted students from a specific level. Students attending low and middle schools are associated to a much lower direct effect compared to the high school students. The coefficient for low and middle school students reveals that additional 30 days of school closures leads to a decrease of about 0.5\% in their Mathematics score. In contrast, the same amount of days of school closures corresponds to a decrease of about 1.8\% in their Mathematics score for high school students. 
It is possible to note also an high degree of heterogeneity in the role of parents' education under school closures. While the parents' educational background continues to moderate the impact of additional days of school closures for low and middle school students, this effect is absent for the specification considering high school students. Indeed, the educational background of the parents plays no role in reducing this impact as the coefficient is not significant. 
Considering the interaction between peers' score and treatment variable, the specification suggests that, again, high school students may have observed the higher loss from school closures. This coefficient is higher for high school students and statistically different from the one of low and middle school students. Also, note that this is in line with what shown in figure \ref{Fig_coefficients}, where under equilibrium, high school students learning appeared to be dependent on peers more than for other students. School closures, therefore, appear to have posed a double burden on high school students, who were unable to compensate the loss in education through their parents, and who have been prevented to benefit from the connection with their peers.  For students attending lower school levels, i.e. low and middle schools, the impact has been much smaller, but remains negative and significant. In contrast with what observed for high school, the level of parents education functions as a compensating mechanism to school closures for students from low and middle schools.



\begin{figure}[h!]
	\begin{center}
		\includegraphics[width=0.60\textwidth]{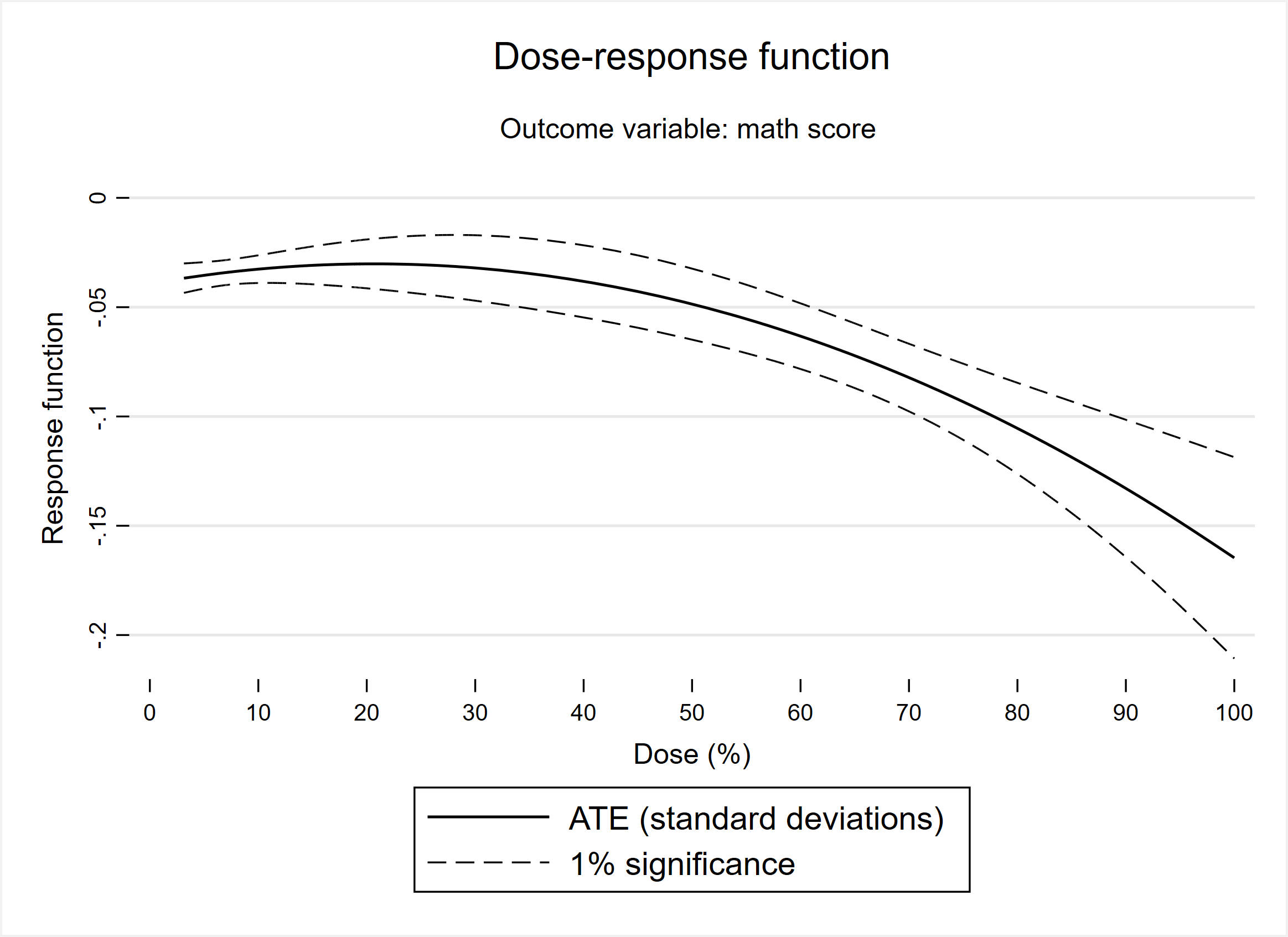}
		\caption{Dose response function of school closures on math score}
		\label{fig:dose_response}
	\end{center}
\end{figure}

Finally, in column 6 of Table \ref{tab:days_closure}, we test whether the results are affected by additional residual endogeneity due to measurement errors. This may derive, for example, by a situation where two students observe the same number of lost days but higher turnover of peers, teachers and staff. Or, alternatively, it can result from differences in timing of days lost, with one student observing 30 days of closures divided over 5 days periods across the school year and another experiencing a unique period of 30 days of school closures. To this scope, we run a 2SLS Instrumental Variable regression on the cross-sectional 2020/2021 sample and we instrument the days of closures with the mean and variance of Covid-19 cases per population. This approach relies on the exclusion restriction that Covid-19 cases will affect the test scores only through school closures. While students may themselves have experienced Covid-19 and loss some distance learning school days, we believe that this effect will be minimal due both to the limited symptoms of Covid-19 on the young population, and to the fact that we consider the average cases at municipal level. We expect that the mean of Covid-19 cases will be positively correlated with the days of school closures, as additional cases on average will push municipalities in keeping the school closed and rely on the distance learning. In contrast, we expect that the variance will be negatively correlated with days of school closures, as additional variance keeping the mean constant means that the cases will be more spread all over the school year. This translates into lower cases per day and thus lower likelihood of school closure. As expected, the mean of Covid-19 cases has a positive effect on the days of closures, while the variance across the year has a negative effect. Both coefficients are significant at 1\% level, the F-test (1293.29) is largely above the rule of thumb, and the test on over-identification does not reject the validity of our instruments (see bottom part of Table \ref{tab:days_closure}). The second stage coefficient linked to the indicator on days of school closures increases substantially. While the magnitude of this result may appear to diverge from the baseline one, it is necessary to acknowledge that this one is calculating the local average treatment effect conditional on Covid-19 cases, while the baseline one can be interpreted as the average treatment effect.
Finally, to take advantage of the continuous nature of the treatment variable, we focus on the cross-sectional 2020/21 sample and run a dose response function with a two degrees polynomial to test whether additional school days may have a non linear effect on the score. This exercise needs the treatment to be distributed between 0-100, thus we re-scale our days of school closures indicator to comply with this criterion. As displayed in Figure \ref{fig:dose_response}, the impact of an additional school days appears almost linear for the low part of the distribution, while an additional day of school closure appears to contribute more to educational losses when the students already observed the 40-50\% of day of school closures in our sample.

\begin{table}[H]
	
	\begin{center}
		\resizebox{0.95 \columnwidth}{!}{
			\begin{threeparttable}
				\caption{Days of school closure and educational score} \label{tab:days_closure}
				\begin{tabular}{lcccccc}
					
					\hline
					
					& \multicolumn{5}{c}{\textbf{Panel}}  & \multicolumn{1}{c}{\textbf{Cross-section} (year=2021)} \\ 
					Model & \multicolumn{3}{c}{\textbf{OLS with FE}}  & Low \& Middle School & High School  & \textbf{2SLS-IV}\\  \hline
					
					VARIABLES & \textbf{Math (std.)} & \textbf{Math (std.)} & \textbf{Italian (std.)} & \multicolumn{3}{c}{\textbf{Math (std.)}} \\
					& (1) & (2) & (3) & (4) & (5) & (6) \\
					\hline
					&  &  &  &  &  &  \\
					\textbf{Days of school closure}  & -0.006*** & -0.060*** & -0.038*** & -0.027*** & -0.083*** & -0.271*** \\
				 & (0.002) & (0.003) & (0.002) & (0.004) & (0.003) & (0.062) \\ \hline

					\textbf{Days of school closure} X \textbf{Peers' score in math test} &  & -0.073*** &  & -0.055*** & -0.080*** &  \\
					 &  & (0.003) &  & (0.004) & (0.003) &  \\
					\textbf{Days of school closure} X \textbf{Years of parents' education} &  & 0.010*** & 0.006*** & 0.021*** & -0.000 &    \\
				     &  & (0.001) & (0.001) & (0.002) & (0.002) & \\
					
					\textbf{Days of school closure} X \textbf{Peers' score in Italian test} &  &  & -0.070*** &  &  &  \\
					&  &  & (0.002) &  &  &  \\ \hline
					
					&  &  &  &  &  &  \\
					\textbf{Other controls} & Yes & Yes & Yes & Yes & Yes & Yes \\
					\textbf{Student FEs} & Yes & Yes & Yes & Yes & No & No \\
					\textbf{Relative time dummies} & Yes & Yes & Yes & No & No & No \\ 
					\textbf{Grade dummies} & Yes & Yes & Yes  & Yes   & Yes  & Yes \\  \hline 
					Observations & 334,366 & 334,366 & 334,366 & 230,332 & 104,034 & 86,140 \\
					R-squared & 0.564 & 0.567 & 0.412 & 0.554 & 0.603 & 0.317 \\ \hline
			
				    & \multicolumn{6}{c}{\textbf{First stage results and statistics}}   \\ \hline
				   
				   Covid-19 cases per person &  &  &  &  &  & 0.028*** \\
				    &  &  &  &  &  & (0.002) \\
				   Variance of Covid-19 cases	&  &  &  &  &  & -0.000*** \\
				    &  &  &  &  &  & (0.000) \\
				    																				
					F-test (P-value) &  &  &  &   &  & 1293.29 (0.000) \\
					Hansen overid. test (P-value) &  &  &   &   &  & 1.482 (0.223) \\ \hline
					
				\end{tabular}
				
				\begin{tablenotes}
					\item \textit{Notes: }{the table reports the estimates of a two-way fixed effect on  panel data (column 1-5)  on students from Sicily, and from a 2SLS model including the total number and the variance of Covid-19 cases as instruments on the 2020/21 cross sectional sample (column 6). Robust standard errors are reported in parentheses. Level of significances ***, **, and * denote significance at 1\%, 5\%, and 10\%, respectively.}{\footnotesize\par}
				\end{tablenotes}
			\end{threeparttable}
			
		}
	\end{center}
\end{table}

\section{Heterogeneity and robustness} \label{hetrob}

This section presents additional heterogeneity results based on some characteristics of production function inputs using the Sicilian case. 
Figure \ref{Fig_coefficients2} below shows the coefficients obtained when pooling the sample across the parents' type of employment, parents' origins, and class size. For parents' employment we consider whether the parents' are unemployed or blue collars, or any combination of these, and we compare these with the remaining employment categories. For parents' origin, we consider the case of both parents' having foreign origins and compare this with cases where at least one parent is Italian. Unfortunately, the INVALSI data do not provide the information about the precise state of origin of the parents, so we are unable to identify whether they were born in a developing or developed country. 

The top-left panel of Figure \ref{Fig_coefficients2} report the effects of days of school closures depending on parents' type of employment and suggest that students experiencing the higher loss are the ones with parents being either unemployed or blue collar. This is in line with the literature showing how the current school system is not filling the role of great leveler missing role of great leveler \citet{agostinelli2022great},  and suggest that school closures may have had a prominent role in increasing inequalities between students, even within the same school and receiving the same inputs. In magnitude terms, the percentage difference of an additional days of school closure for these disadvantaged students is 35\% higher than the rest of the sample.

The top-right panel of Figure \ref{Fig_coefficients2} reports the coefficients obtained when pooling along the parents' origins. We observe higher variance for the coefficients of students with foreigner parents, but the two coefficients are not statistically different and the confidence intervals are overlapping. The higher variance likely derives by the fact that our sample includes both students from developed and developing countries, with likely the first experiencing a lower impact than the second, but this cannot be disentangled with the available data. 

The bottom panel of Figure \ref{Fig_coefficients2} reports the estimated coefficients obtained when pooling the sample across class sizes below and above the regional median. The results suggest that the impact of an additional day of school closure changes substantially across this dimension, with students attending small classes experiencing a loss that is about 50\% higher than the ones attending classes larger than the median. There may be at least two explanations for this finding. First, as suggested by the development or education economics literature \citet{case1999school}, the teacher/pupil ratio is one of the main determinants of the human capital accumulation during schools as it ensure higher quality of teaching inputs. Therefore, for an additional day of school closures, the students attending smaller size classes are losing higher quality input and experience an higher loss. Second, the socio-economic literature stresses that in smaller groups, the social ties are stronger and individuals are more influenced by their peers. For example, the network literature \citet{mcpherson2006social} show that, in USA, from 1985 to 2004 there is a reduction in number of confidants within discussion core groups, together with a higher network density, defined as the mean intensity of tie strength among the discussion partners mentioned. Our result, therefore, could derive by the fact that loosing the direct contact with the peers, in groups where these ties are stronger, may imply higher costs in terms of educational score.

In Table \ref{tab:rob1} below, we report some robustness checks to take into account how choices on variable specification and group comparison may potentially drive our results. In particular, in the first three columns we report the results obtained when using another transformation of the dependent and explanatory variables. In column 1 we take the natural log of the continuous dependent and explanatory variables, adding 1 to include the zero valued observations. In column 2 we use the Inverse Hyperbolic Sine Transformation (IHS) developed by \citet{bellemare2020elasticities}. Both these exercises show very similar results and the magnitude of the coefficients, when duly accounting for the difference in the functional form, is similar to the baseline. In column 3 we report the results when taking the level of the dependent and explanatory variables. Again, the results remain consistent to this additional exercise. Finally, in column 4, we report the results obtained when adding inverse probability weight to the specification of column 1 in Table \ref{tab:days_closure}. These weights derives from scores obtained by running a propensity score matching on the probability of being treated and controlling for a large set of covariates, including  Mathematic and Italian score at baseline, parents' year of education, average peers' score at baseline, province dummies, parents' employment typology, number of classes and total population. As shown in Table \ref{tab:rob1}, also this additional test confirms the negative impact of experiencing additional days of school closures on the test score. 
Finally, Table \ref{tab:rob_national} in the Appendix reports the results from a similar set of robustness test for the national level data. In particular, we run the three specifications taking the natural log of the dependent and explanatory variables, the Rasch score and level variables, and the inverse probability weighted standardized variables. Again, the results remain consistent with the benchmark.

\begin{figure}[H]
	\centering
	\minipage{0.5\textwidth} 
	\includegraphics[width=1\linewidth]{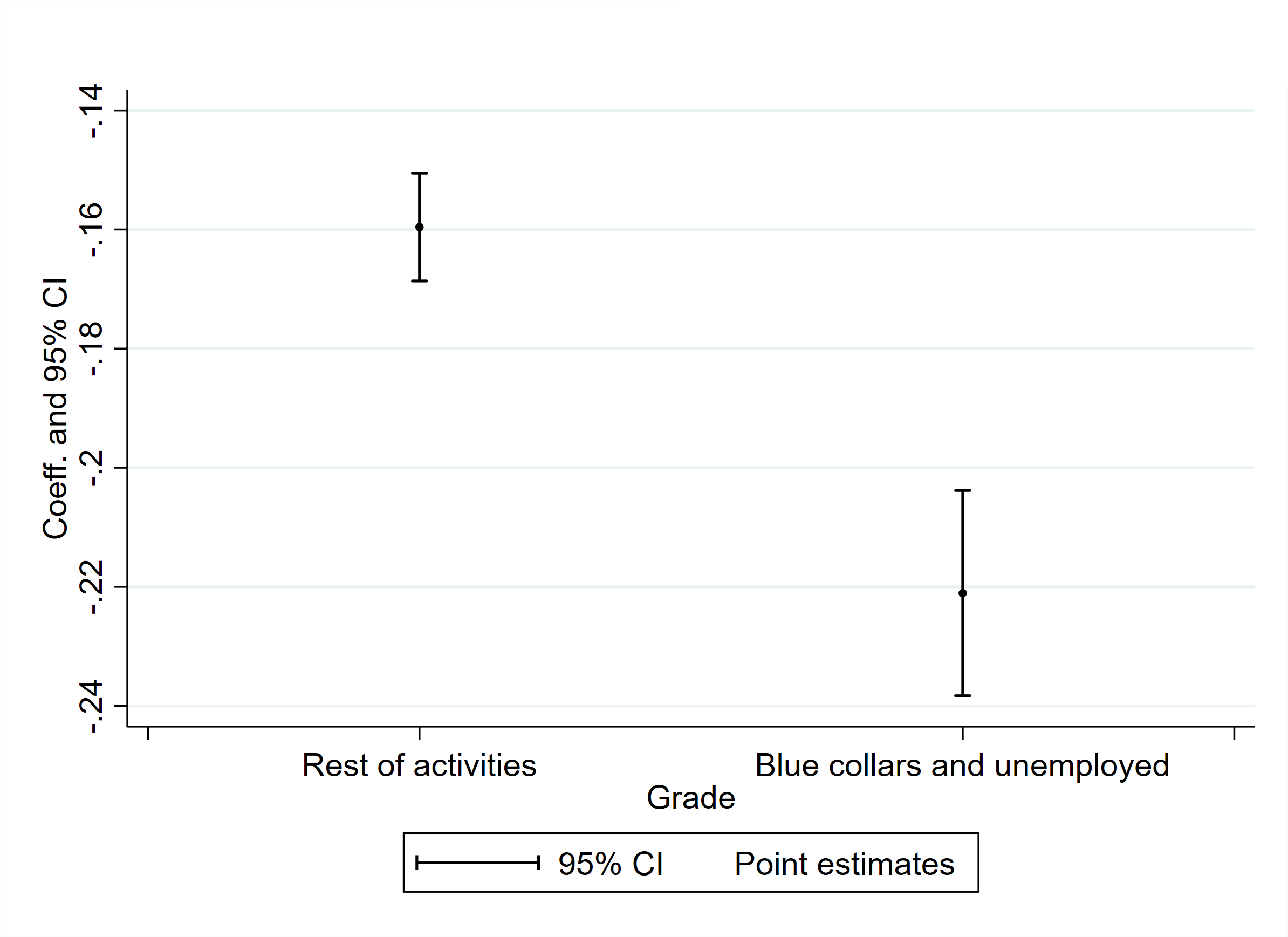}
	\endminipage\hfill
	\minipage{0.5\textwidth}
	\includegraphics[width=1\linewidth]{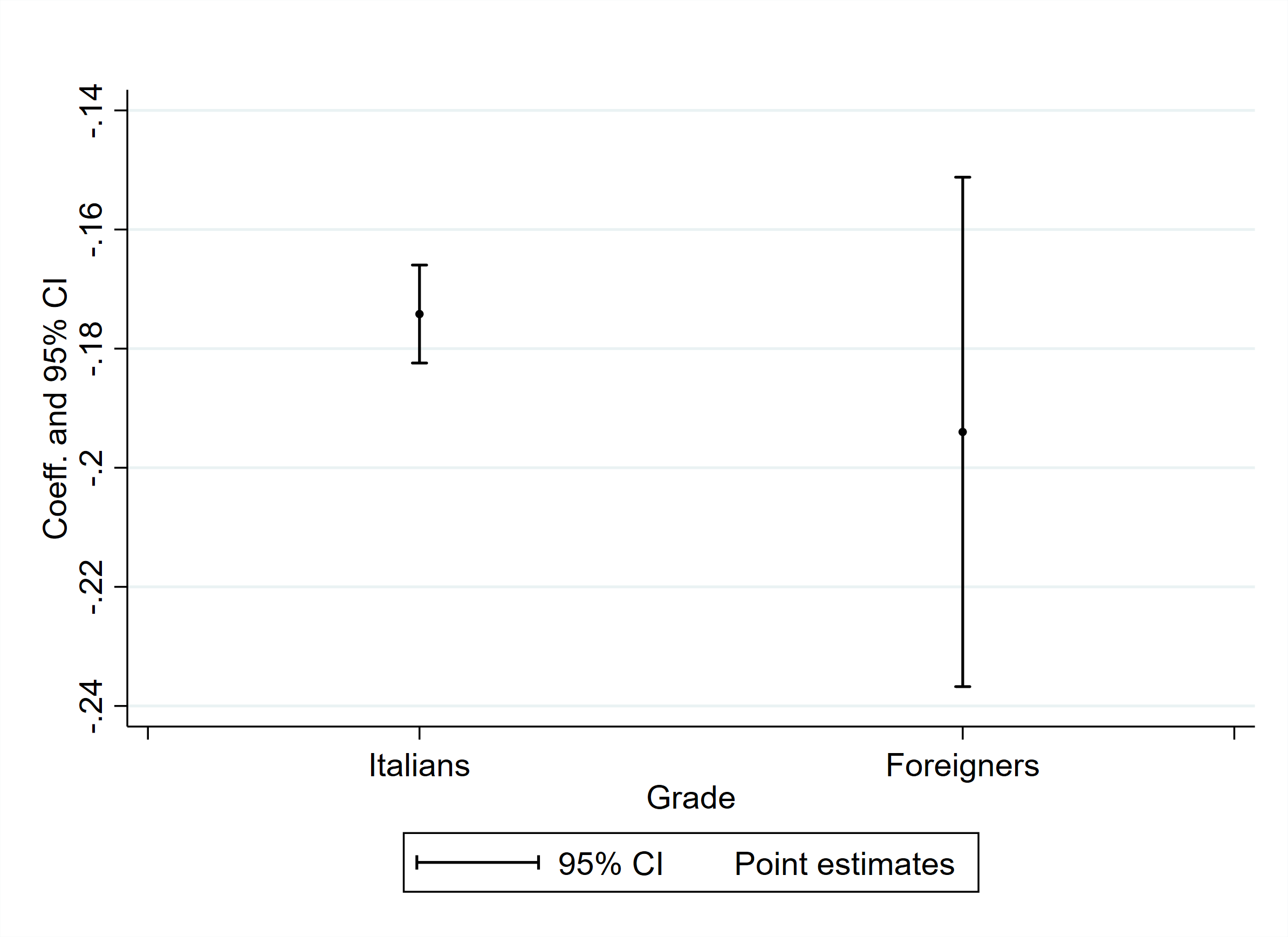}
	\endminipage\hfill
	\minipage{0.5\textwidth}
	\includegraphics[width=1\linewidth]{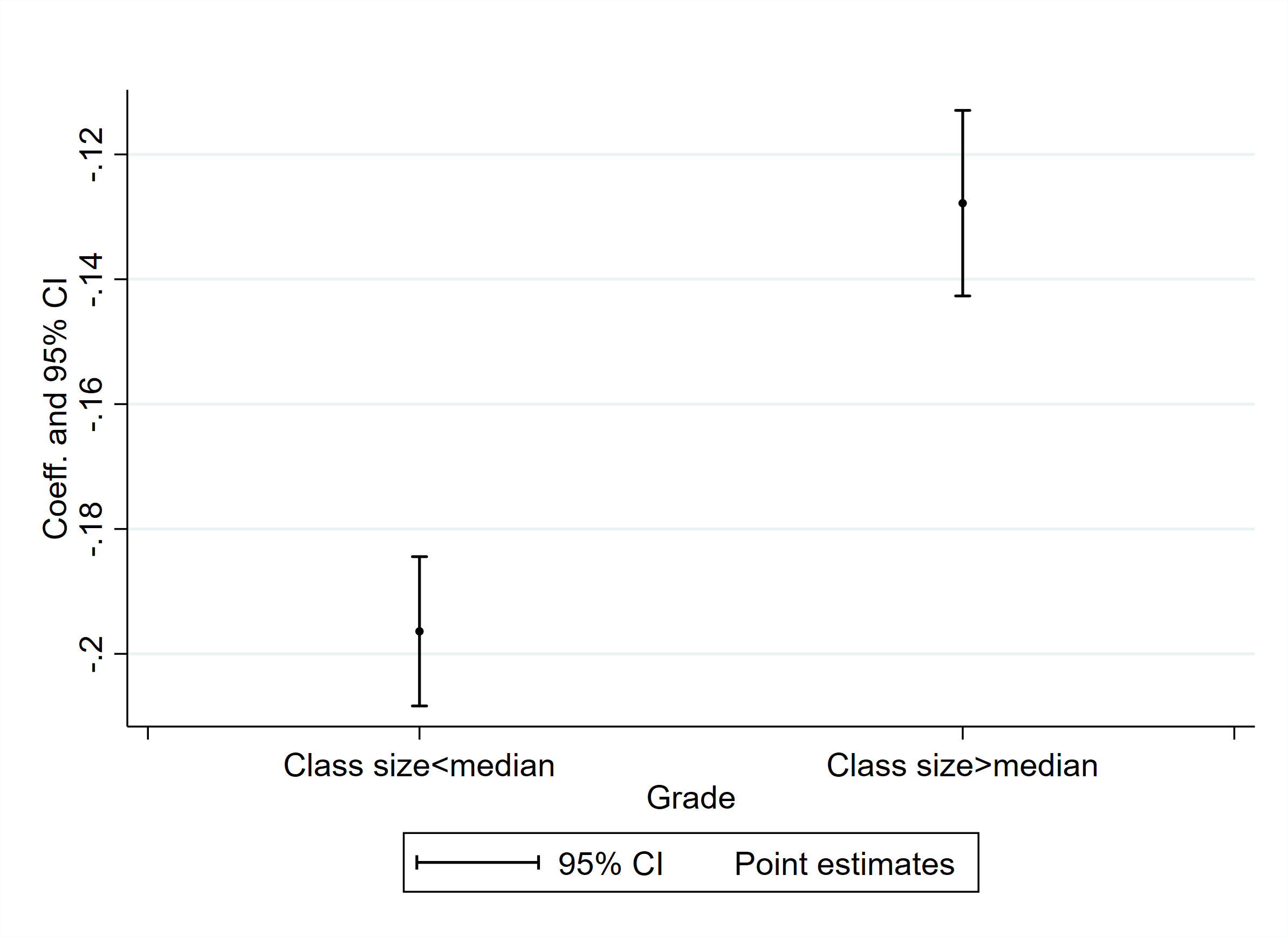}
	\endminipage\hfill
	\caption{Heterogeneity by parent's professions, origins and class size}. 
	\label{Fig_coefficients2}
\end{figure}

\begin{table}[H]
	
	\begin{center}
		\resizebox{0.95 \columnwidth}{!}{
			\begin{threeparttable}
				\caption{Robustness specifications on the effect of school closures on test scores in Sicily} \label{tab:rob1}
				\begin{tabular}{lcccc}
					
					\hline

					& Math (ln) & Math (IHS) & Math  (Rasch) & Math (std. - PSM)   \\ 
					
					& (1) & (2) & (3) & (4)  \\
				\hline
				
				\textbf{Days of school closure} (ln) & -0.001*** & &  &  \\
				& (0.000) &  &  &  \\ 
				\textbf{Days of school closure} (IHS) &  & -0.001*** &  &  \\
			&  & (0.000) &  &  \\ 
			
			\textbf{Days of school closure} (level) &  & &  -0.006*** &  \\
		   &  &  & (0.002) &   \\

		\textbf{Days of school closure} (std.) &  &  &  & -0.006***   \\
	&  &  &  & (0.002)  \\
	&  &  &  &  \\ \hline
	\textbf{Other controls} & Yes & Yes & Yes & Yes  \\
	\textbf{Student FEs} & Yes & Yes & Yes & Yes  \\
	\textbf{Relative time dummies} & Yes & Yes & Yes & Yes  \\ 
	\textbf{Grade dummies}& Yes & Yes & Yes & Yes \\  \hline 
	
	\textbf{Observations} &  334,366 & 334,366 &  334,366 & 334,366  \\
	R-squared & 0.553 & 0.554 & 0.563 & 0.534   \\
	\textbf{Number of students} & 167,183  & 167,183  & 167,183  & 167,183   \\ \hline
	
				\end{tabular}
				
				\begin{tablenotes}
					\item \textit{Notes: }{the table reports the estimates from a Two-way fixed effect model on the impact of Covid-19 related closures on test scores. The main explanatory variable is a variable capturing the number of days of school closure. he specification corresponds to the one reported on column 2 of Table \ref{tab:days_closure}. Columns 1 considers the natural log of the dependent and explanatory continuous variables. Columns 2 reports the results when using the Inverse Hyperbolic Sine Transformation developed by \citet{bellemare2020elasticities}. Column 3 displays the results with the level variables. Column 4 is similar to the baseline but includes inverse probability weights calculated through a propensity score matching (PSM). PSM includes the treatment as dependent variable and, as controls, the following variables: score in Mathematic and Italian at baseline, year of educations of parents at baseline, average peers' score at baseline, province dummies and parents' employment typology, number of classes and total population. Robust standard errors in parentheses, level of significances ***, **, and * denote significance at 1\%, 5\%, and 10\%, respectively.}{\footnotesize\par}
				\end{tablenotes}
			\end{threeparttable}
			
		}
	\end{center}
\end{table}

\section{Conclusions} \label{conclusion}

In the last two years, Governments strongly relied on school closures to reduce the diffusion of Covid-19 , especially during the first waves \citep{haug2020ranking,Hsiang2020covid,kucharski2020effectiveness}. While a growing literature discussed the potential effects in terms of several dimensions of future human capital losses, most of this literature used theoretical assumptions of standard human capital models and a common shock on the present cohort with respect to the previous ones that did not experience these closures.

This work participated to the debate, by using a newly collected dataset of local closures by schools (and Covid-19 cases at local level), motivated by the fact that after the first months of general lockdowns, closures were not equally distributed but depending by virus diffusion criteria on local indicators.  While at a general level we found that generational losses are comparable with what literature found, the within estimation shows as 30 additional days of closures implied a 1\% higher annual drops in educational score implemented by the general Italian agency of students evaluation. The impact is higher for high school students (1.8\%) than for low and middle school ones (0.5\%). This is likely explained by the the fact that the educational production function of younger students depend more on their parents and by the higher parental ability to cope with the shock when children are younger. The findings also show a more pronounced impact on more disadvantaged students, which has likely translated in more unequal educational attainment and human capital formation across the social stratification in Italy.

While the present work has estimated the short-term educational costs of the so-called Covid-19 mitigation policy on school closures, it leaves open a set of questions about how the long-term perspective of the affected students will look like. Whether these short-term costs will translate in long-term lower salaries, as some work predicts, or if any policy actions will impede the expected growing inequality, is left to future work.


\clearpage

 \bibliography{Invalsi3.bib}
\clearpage

\appendix

\begin{table}[H]
	
	\begin{center}
		\resizebox{0.95 \columnwidth}{!}{
			\begin{threeparttable}
				\caption{Robustness specifications on the effect of school closures on test scores at national level} \label{tab:rob_national}
				\begin{tabular}{lcccccc}
					
					\hline

					& Math (ln) & Italian (ln) & Math  (Rasch) & Italian (Rasch) & Math (std. - PSM) & Italian (std. - PSM)  \\ 
					
					& (1) & (2) & (3) & (4) & (5) & (6) \\
					\hline
					
					\textbf{School closures} & -0.011*** & -0.003*** & -1.569*** & -0.434*** & -0.017*** & -0.006*** \\
					(Treated=1 X t=1) 	& (0.000) & (0.000) & (0.041) & (0.042) & (0.001) & (0.001)  \\ \hline
					
					\textbf{Other controls} & Yes & Yes & Yes & Yes & Yes & Yes \\
					\textbf{Student FEs} & Yes & Yes & Yes & Yes & Yes & Yes \\
					\textbf{Relative time dummies} & Yes & Yes & Yes & Yes & Yes & Yes \\ 
					\textbf{Grade dummies}& Yes & Yes & Yes & Yes & Yes & Yes \\  \hline 
					
					\textbf{Observations} & 4,613,714 & 4,613,714 &  4,613,714 & 4,613,714 & 4,613,714 & 4,613,714  \\
					R-squared & 0.395 & 0.274 & 0.413 & 0.280 & 0.379 & 0.261 \\
					\textbf{Number of students} & 2,306,857 & 2,306,857 &  2,306,857 & 2,306,857  & 2,306,857 & 2,306,857  \\ \hline
					
				\end{tabular}
				
				\begin{tablenotes}
					\item \textit{Notes: }{the table reports the estimates from a Two-way fixed effect model on the impact of Covid-19 related closures on test scores. The main explanatory variable is a dummy activating for the treated group after the Covid-19 pandemic occurred. The specification corresponds to the one reported on column 2 of Table \ref{tab:result1}. Columns 1-2 consider the natural log of the dependent variables and explanatory continuous variables. Columns 3-4 reports the results when using the level variables from the Rausch model reported by INVALSI. Columns 5-6 display the results from the baseline specification with standardized variables but includes inverse probability weights calculated through a propensity score matching (PSM). PSM includes the treatment as dependent variable and, as controls, the following variables: score in Mathematic and Italian at baseline, year of educations of parents at baseline, average peers' score at baseline, province dummies and parents' employment typology. Robust standard errors in parentheses, level of significances ***, **, and * denote significance at 1\%, 5\%, and 10\%, respectively.}{\footnotesize\par}
				\end{tablenotes}
			\end{threeparttable}
			
		}
	\end{center}
\end{table}

\end{document}